\documentclass[usenatbib]{mn2e}
\usepackage{enumitem}
\usepackage{amsmath}
\usepackage{graphicx}
\usepackage{epsfig}
\usepackage{url}                                 
\usepackage{morefloats}  
\usepackage{amssymb}
\usepackage{rotating}
\usepackage{natbib}
\usepackage{lineno}
\usepackage{rotfloat}
\usepackage{threeparttable}
\usepackage{textcomp}
\usepackage{longtable}
\usepackage{multicol}
\usepackage{paralist}
\usepackage{amsfonts}

\title[ETG in Different Environments]{H-ATLAS/GAMA and HeViCS - Dusty Early-Type Galaxies in Different Environments}
\author[N.~K.~Agius et~al.]{\parbox{\textwidth}{N.~K.~Agius,$^{1}$
S.~di~Serego~Alighieri,$^{2}$
S.~Viaene,$^{3}$
M.~Baes,$^{3}$
A.~E.~Sansom,$^{1}$\thanks{E-mail: \texttt{AESansom@uclan.ac.uk}}
N.~Bourne,$^{15}$
J.~Bland-Hawthorn,$^{4}$
S.~Brough,$^{5}$
T.~A.~Davis,$^{6}$
I.~De~Looze,$^{3,17}$
S.~P.~Driver,$^{7,8}$
L.~Dunne,$^{9,15}$
S.~Dye,$^{16}$
S.~A.~Eales,$^{13}$
T.~M.~Hughes,$^{3}$
R.~J.~Ivison,$^{6,15}$
L.~S.~Kelvin,$^{7,8,10}$
S.~Maddox,$^{16}$
S.~Mahajan,$^{11}$
C.~Pappalardo,$^{12}$
A.~S.~G.~Robotham,$^{7,8}$
K.~Rowlands,$^{8}$
P.~Temi$^{14}$
E.~Valiante$^{13}$}
\vspace{0.4cm}
\\
\parbox{\textwidth}{$^{1}$Jeremiah Horrocks Institute, University of Central Lancashire, Preston, PR1 2HE, Lancashire, UK\\
$^{2}$INAF - Osservatorio Astrofisico di Arcetri, 50125 Firenze, Italy\\
$^{3}$Sterrunkundig Observatorium UGent, Krijgslaan 281 S9, B-9000 Gent, Belgium.\\
$^{4}$Sydney Institute for Astronomy, School of Physics A28, University of Sydney, NSW 2006, Australia.\\
$^{5}$Australian Astronomical Observatory, PO Box 915, North Ryde, NSW 1670, Australia.\\
$^{6}$European Southern Observatory, Karl-Schwarzschild-Str. 2, 85748 Garching, Germany.\\
$^{7}$International Centre for Radio Astronomy Research, The University of Western Australia, 35 Stirling Highway, Crawley, WA 6009, Australia.\\
$^{8}$School of Physics and Astronomy, University of St Andrews, North Haugh, St Andrews, Fife, KY16 9SS, UK.\\
$^{9}$Department of Physics and Astronomy, University of Canterbury, Private Bag 4800, Christchurch, 8140, New Zealand.\\
$^{10}$Institut für Astro- und Teilchenphysik, Universität Innsbruck, Technikerstraße 25, 6020 Innsbruck, Austria.\\
$^{11}$School of Mathematics and Physics, University of Queensland, Brisbane, QLD 4072, Australia.\\
$^{12}$CAAUL, Observatorio Astronomico de Lisboa, Universidade de Lisboa, Tapada de Ajuda, 1349-018, Lisboa, Portugal.\\
$^{13}$School of Physics and Astronomy, Cardiff University, The Parade, Cardiff, CF24 3AA, UK.\\
$^{14}$Astrophysics Branch, NASA/Ames Research Center, MS 245-6, Moffett Field, CA 94035.\\
$^{15}$Institute for Astronomy, University of Edinburgh, Royal Observatory, Blackford Hill, Edinburgh EH9 3HJ, UK. \\
$^{16}$School of Physics and Astronomy, University of Nottingham, Nottingham, NG7 2RD, UK. \\
$^{17}$Institute of Astronomy, University of Cambridge, Madingley Road, Cambridge, CB3 0HA, UK. 
}}

\begin{document}

\pubyear{2013} \volume{000} \pagerange{\pageref{firstpage}--\pageref{lastpage}} \pubyear{2013}

\maketitle
\label{firstpage}

\begin{abstract}

The \textit{Herschel Space Observatory} has had a tremendous impact on the study of extragalactic dust. Specifically, early-type galaxies (ETG) have been the focus of several studies. In this paper we combine results from two $Herschel$ studies - a Virgo cluster study HeViCS and a broader, low-redshift H-ATLAS/GAMA study - and contrast the dust and associated properties for similar mass galaxies. This comparison is motivated by differences in results exhibited between multiple $Herschel$ studies of early-type galaxies. A comparison between consistent modified blackbody derived dust mass is carried out, revealing strong differences between the two samples in both dust mass and dust-to-stellar mass ratio. In particular, the HeViCS sample lacks massive ETG with as high a specific dust content as found in H-ATLAS. This is most likely connected with the difference in environment for the two samples. We calculate nearest neighbour environment densities in a consistent way, showing that H-ATLAS ETG occupy sparser regions of the local Universe, whereas HeViCS ETG occupy dense regions. This is also true for ETG that are not $Herschel$-detected but are in the Virgo and GAMA parent samples. Spectral energy distributions are fit to the panchromatic data. From these we find that in H-ATLAS the specific star formation rate anticorrelates with stellar mass and reaches values as high as in our Galaxy. On the other hand HeViCS ETG appear to have little star formation. Based on the trends found here, H-ATLAS ETG are thought to have more extended star formation histories and a younger stellar population than HeViCS ETG. 
\end{abstract}


\section{Introduction}

Dust is a fundamental component of the interstellar medium (ISM) of galaxies for the thermodynamics and chemistry of the gas, for the dynamics of the accretion in dense star-forming clouds, and for the attenuation of UV/blue radiation and its re-emission in the far infra-red (FIR; \citealp{draine_2003}). The relative amount of dust varies strongly with galaxy type, increasing by about three orders of magnitude on average along the Hubble sequence (e.g. \citealp{cortese_2012}). The connection between dust and chemical evolution also varies with galaxy type. In late-type galaxies (LTG) dust is strongly linked with star formation (SF), both because it serves as a catalyst for the formation of molecular gas necessary for SF and because, being heated mostly by young stars, its emission traces the regions of SF. The same paradigm does not necessarily apply to early-type galaxies (ETG; comprising of ellipticals and lenticulars), where dust can be heated by the radiation field produced by evolved stars and can be more diffuse, therefore not serving as a SF catalyst. In addition, ETG, particularly those in clusters, can have much larger amounts of hot gas than LTG, not favouring the presence of dust.

It is therefore important to study dust in ETG separately from LTG and the \textit{Herschel Space Observatory} \citep{pilbratt_2010} has allowed several such detailed studies (\citealp{rowlands_2012,smith_HRS,sperello_2013}, S13 hereafter, and \citealp{agius_2013}, A13 hereafter). In particular S13 searched for dust in a large optical sample of 910 ETG in the Virgo cluster, extending also to dwarf ETG, using the PACS \citep{poglitsch_photodetector_2010} and SPIRE \citep{griffin_herschel-spire_2010} instruments, and found it in 17$\%$ of the elliptical galaxies, in 40$\%$ of lenticulars (S0 + S0a) and in about 3$\%$ of the dwarfs (dE + dS0). They showed that the presence of dust does not correlate with the presence of neutral gas (HI) and the dusty ETG do not appear to have bluer B-H colours, i.e. to be more star-forming than the non-dusty ones. On the other hand A13 searched for dust, also with PACS and SPIRE, in a sample of 771 brighter ETG (M$_r <-17.4$mag) over a very large volume (144 deg$^{2}$ and 0.013$\le$z$\le$0.06) and found it in 20$\%$ of the ellipticals and in 38$\%$ of the lenticulars. Not only are these detection rates high, but also the relative amount of dust is higher than in HeViCS and the dusty ETG have bluer colours, suggesting that they may be forming more stars. Furthermore S13 found a dependence of the dust temperature on the stellar mass and on the average B-band surface brightness within the effective radius, but A13 did not.

These differences may be an effect of the environment, since H-ATLAS covers a wide range of environments, potentially hosting extreme mergers or interactions, whilst HeViCS is limited to a cluster environment, which may be less favourable for dust. 
\citet{bourne_2012} pointed out that environment was a possible influence on the relatively low dust masses in Virgo cluster ETG, in comparison with their results for stacked red sequence galaxies from the GAMA survey.
Alternatively the dusty H-ATLAS ETG could represent younger versions of the `standard' ETG; i.e. they may have formed recently, or have more extended star formation histories. In fact, if galaxy evolution is influenced by the environment, ETG in sparse environments are more likely to be at an earlier stage in their evolution than ETG in dense environments (e.g. \citealp{thomas_2005}). Some differences may also be explained by the different models used to estimate the dust mass or by limitations in the models or wavebands used. Although a single modified blackbody (modBB) fitting approach gives a good estimate of the mass of cold, diffuse dust grains in the ISM of galaxies at all redshifts \citep{dunne_2000,blain_2003,pope_2006,dye_2010,bianchi_2013}, it may not account for the emission from dust in warmer media, such as the grains surrounding birth clouds of hot, young stars, or colder dust that needs multiple temperature models. The addition of further modBBs allowing for different temperatures have been shown to improve such fits \citep{dunne_2001,galametz_2011,dale_2012}. Furthermore, given the wealth of panchromatic data for HeViCS and H-ATLAS/GAMA ETG, it is possible to exploit multi-wavelength SED fits which consider stellar emission at UV/optical wavelengths, the attenuation by dust and resultant emission in the infrared.

Temi et al. (2009a,b) and Amblard et al. (2013) have further investigated the diversity of ETG by studying the physical properties of a sample of local E and S0 galaxies. They find that many local S0 galaxies are quite distinct from ellipticals, containing dust and cold gas in amounts that may be sufficient to generate appreciable star formation at rates as large as several M$_{\odot}$/yr. However in this paper we cannot investigate in detail the differences between E and S0 galaxies, since they are difficult to distinguish in the H-ATLAS/GAMA sample because of their distance and the limited spatial resolution of SDSS images.

In this paper we compare the properties of the dust in ETG from both the HeViCS sample of S13 and the H-ATLAS/GAMA sample of A13, by characterising them in a uniform way and taking into account data at shorter wavelengths, in an attempt to understand the differences and the reasons causing them. 

This paper is laid out in the following manner. In Section \ref{sec:section2} we describe the data available for the HeViCS and the H-ATLAS/GAMA ETG samples. In Section \ref{sec:section3} we compare the two samples, while in Section \ref{sec:section4} we discuss effects of environment. Panchromatic fits to the ETG SED are described in Section \ref{sec:section5} and a discussion of 
derived parameters is given in Section \ref{sec:section6},
and is followed by our conclusions in Section 7. The photometric data obtained from WISE are described in Appendix A. We assume a flat Universe with $\Omega_{M}$ = 0.3, $\Omega_{\Lambda}$ = 0.7 and H$_{o}$ = 70 km/s/Mpc.

\section{Data}\label{sec:section2}

This section summarises the different surveys utilised in constructing the ETG samples that will be compared in this work.

\subsection{Far-Infrared Surveys}

The Herschel-Astrophysical Terahertz Large Area Survey (H-ATLAS\footnote{http://www.h-atlas.org}) is the widest open-time extragalactic survey with the Herschel Space Observatory, with one of its ultimate science aims being the investigation of the dust content of the nearby Universe at z$<$0.5 \citep{Eales_2010,dunne_2011}. This survey samples over $\sim$570 deg$^{2}$ of sky, covering a range of environments in a uniform way. Their data collection process involved parallel imaging with Herschel's two photometers, PACS (100$\mu$m and 160$\mu$m; \citealp{poglitsch_photodetector_2010}) and SPIRE (250$\mu$m, 350$\mu$m and 500$\mu$m; \citealp{griffin_herschel-spire_2010}), with a 60$^{\prime\prime}$s$^{-1}$ scan rate. H-ATLAS has a 5$\sigma$ sensitivity limit of 33.5~mJy/beam at 250$\mu$m \citep{rigby_2011} - this corresponds to a dust mass of $\sim$10$^{5-7}$M$_{\odot}$ for a range of temperatures (15-30 K) at low redshift (z$\le$0.06).

H-ATLAS catalogues were constructed from maps as described in \citet{pascale_2011} and \citet{ibar_2010}.
Source extraction was based on emission greater than 5$\sigma$ in any of the 3 SPIRE wavebands, described in detail for the Science Demonstration Phase in \citet{rigby_2011}.
\citet{smith_2011} gives a description of the likelihood-ratio analysis performed to identify $r$-band optical counterparts to the SPIRE sub-mm selected sources. Based on the resultant positional and photometric information for the individual sources, PACS flux densities are measured using circular apertures placed at the SPIRE positions. Details of the Phase 1 H-ATLAS and the GAMA catalogues used in this paper can be found in A13.

The Herschel Virgo Cluster Survey (HeViCS\footnote{http://wiki.arcetri.astro.it/bin/view/HeViCS/WebHome}; \citealp{davies_herschel_2010,davies_2012}) is an audit of a large fraction (84 square degrees) of the Virgo Cluster in the same five Herschel bands as the H-ATLAS survey. This specifically samples the dense environment of a nearby cluster, going down to fainter luminosities than H-ATLAS. Additionally their observations are deeper than the H-ATLAS observations, with four linked cross-scans for HeViCS compared to a single cross-scan for H-ATLAS. HeViCS observations were performed in fast-parallel mode with PACS and SPIRE, with a scan rate of 60$^{\prime\prime}$s$^{-1}$. The HeViCS 5$\sigma$ sensitivity at 250$\mu$m is 25-33~mJy for sources smaller than the PSF (\citealp{auld_2013}; S13); depending on the dust temperature this corresponds to a dust mass of $\sim$0.2-1$\times$10$^{5}$M$_{\odot}$ at the 17 Mpc distance of the main Virgo cluster cloud. A detailed account of the data collection, reduction and flux measurements can be found in \citet{auld_2013} and in S13.

\subsection{Multi-Waveband Data}

The Galaxy and Mass Assembly (GAMA\footnote{http://www.gama-survey.org}) is a spectroscopic and photometric survey dedicated to constructing a galactic database which spans the electromagnetic spectrum from ultraviolet to radio wavebands \citep{driver_2011}. This campaign is being supplemented by imaging from surveys such as the Sloan Digital Sky Survey in the optical \citep{abazajian_2009}, GALEX in the UV \citep{bianchi_1999}, UKIDSS-LAS in the NIR \citep{lawrence_2007} and H-ATLAS \citep{Eales_2010} in the FIR/sub-mm; all these surveys have overlapping data within the same regions and the photometry has been made self-consistent using an aperture matching technique described in \citet{hill_2011}.

The spectroscopic element of GAMA has just been completed at the Anglo-Australian Telescope (AAT), with the most recent Data Release (DR2) from GAMA giving access to 70,000 new redshifts in the GAMA I regions \citep{hopkins_2012,liske_2014}. These are three regions of 48 square degrees each, centred at 9, 12 and 14.5 hours (G09, G12 and G15) on the celestial equator. Spectroscopic completeness limits are given as apparent petrosian magnitude r$_{pet}<$19.4 mag in the G09 and G15 fields, and  r$_{pet}<$19.8 mag in G12 for GAMA I.

The H-ATLAS equatorial fields coincide with those of GAMA, and matching between the two sets of data revealed 
$\sim$10,000 
counterparts, using the likelihood ratio method (Bourne et al. in prep). It is from within these counterparts that the H-ATLAS ETG sample is constructed, as described in A13 and $\S$\ref{sec:SubS}. Section 2 of A13 gives details of the GAMAI databases used in this current paper.

A very large set of data is available for galaxies in the Virgo cluster. In this nearby cluster we can obtain accurate galaxy morphological classifications and observe galaxies covering a wide range of luminosities. The main original source of information is the Virgo Cluster Catalogue (VCC, \citealp{binggeli_1985,binggeli_1993}) which, together with Virgo SDSS data \citep{davies_2014}, remains the most complete optical catalogue, until the catalogue of the Next Generation Virgo Cluster Survey (NGVS, \citealp{ferrarese_2012}) becomes publicly available. The VCC is complete to photographic magnitude m$_{pg}$ = 18.0, but also contains fainter galaxies. GOLDMine \citep{gavazzi_2003} provides a compilation of data on VCC galaxies. Useful additions are the GALEX Ultraviolet Virgo Cluster Survey papers (GUViCS, \citealp{boselli_2011,boselli_2012}) and the HI survey for ETG of \citet{serego_2007}.

\section{Overview of ETG Samples}\label{sec:section3}

This section compares and contrasts two ETG samples, which are described below. Particular emphasis is placed on the differences between the classification criteria for these samples. A summary of the results from their parent papers is also given.

\subsection{H-ATLAS Sample}\label{sec:SubS}

From the sample of 771 ETG with M$_{r}\le$-17.4 in the GAMA equatorial fields, the H-ATLAS detections comprise 220 Es and S0s with 5$\sigma$ 250$\mu$m from $Herschel$. Thus there are 551 ETG in our GAMA sample that are below this detection limit (the undetected H-ATLAS ETG). The optical counterparts in GAMA have reliability of association $>0.8$. The morphological classification process for all these ETG is fully described in A13; briefly, it was based on visual classification of blue, green and red optical galaxy cutouts into six groupings of E, S0, SB0a, Sbc, SBbc and Sd galaxies (see \citealp{kelvin_2014} for a full account of this process). The galaxies classified in this way are GAMA I galaxies within a redshift range of 0.013$\le$z$\le$0.06 and complete to an absolute magnitude cutoff of M$_{r}\le$-17.4 - these limits are therefore also applicable to the ETG sample. We note that this sample lacks the faintest ETG close to the M$_{r}$=-17.4 limit, probably because these are dominated by the PSF and therefore excluded since they cannot be reliably classified (see above). We will discuss the effects of this selection in $\S$~\ref{sec:comparison}. 

\begin{figure*}
\begin{center}$
\begin{array}{ccc}
      \includegraphics[width=0.3\textwidth,height=0.3\textwidth]{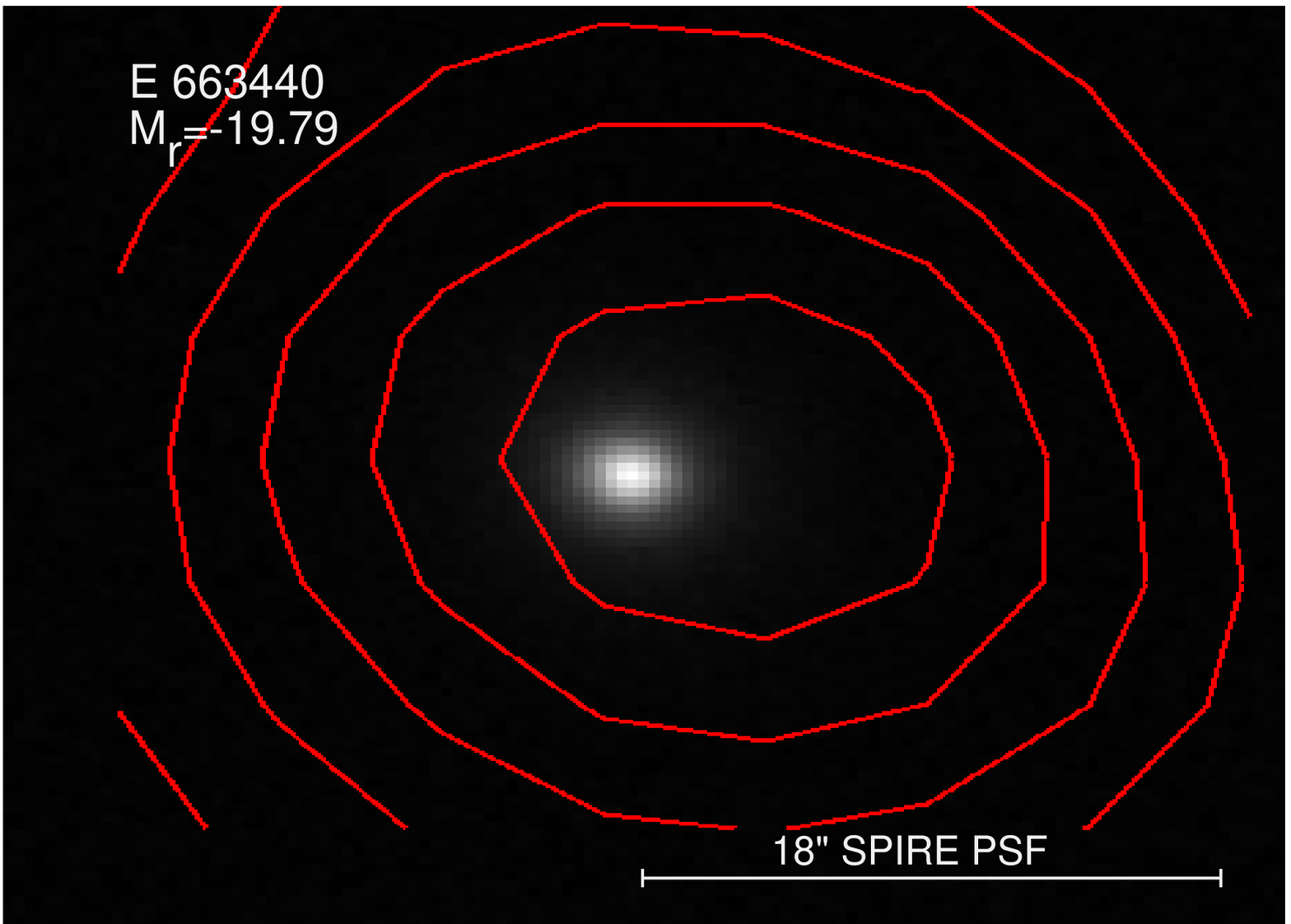} &
   \includegraphics[width=0.3\textwidth,height=0.3\textwidth]{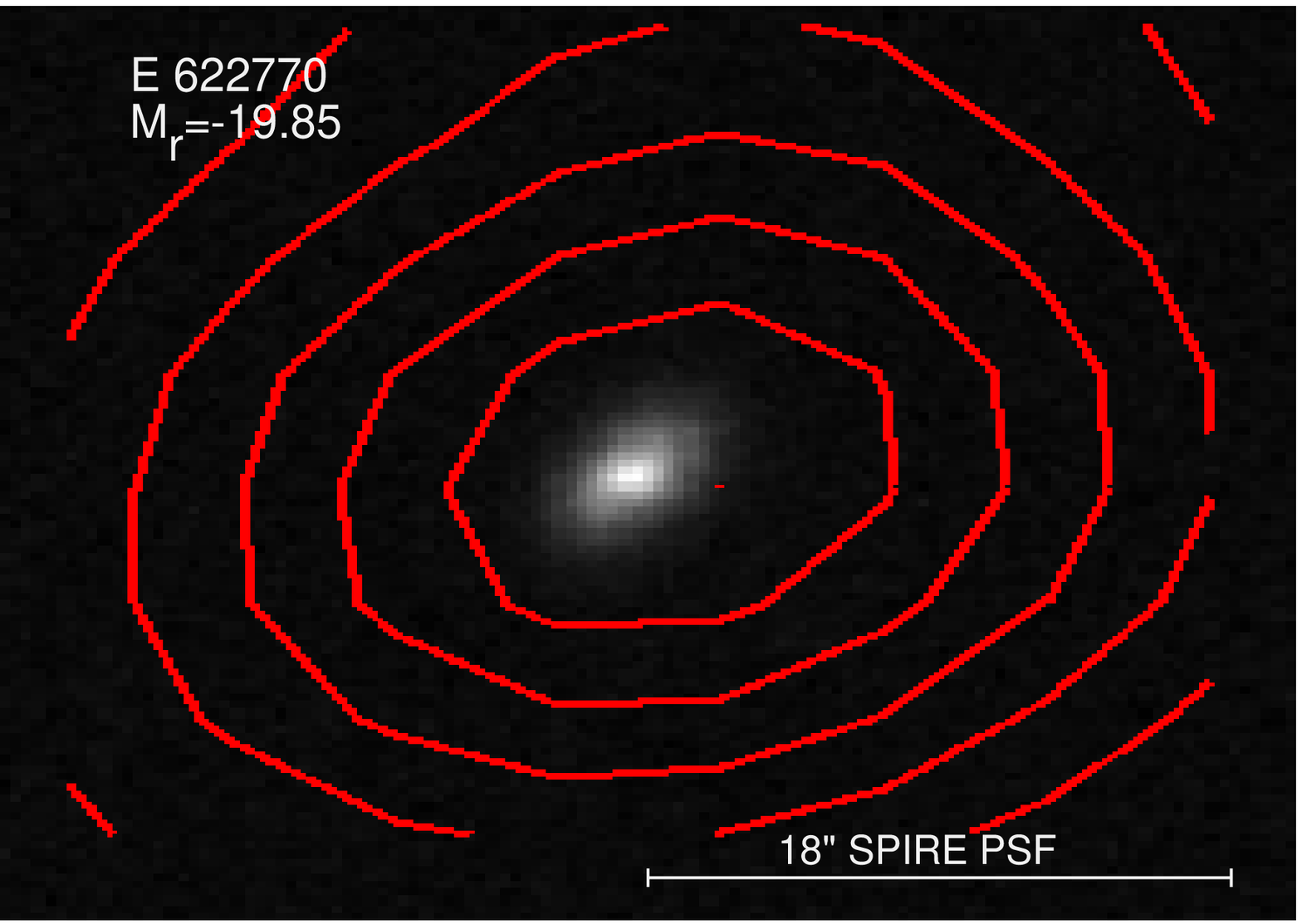} &
    \includegraphics[width=0.3\textwidth,height=0.3\textwidth]{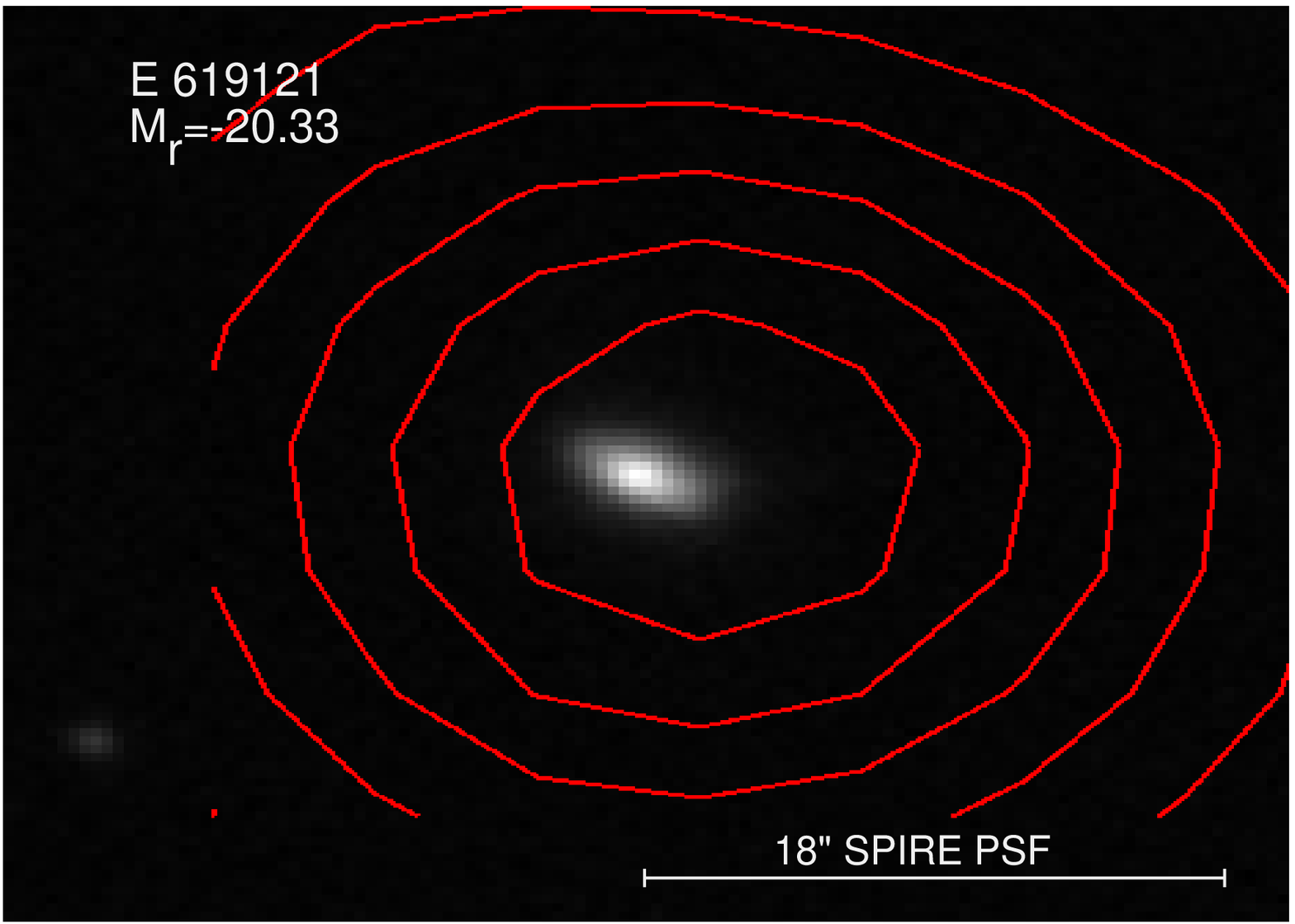} \\
      \includegraphics[width=0.3\textwidth,height=0.3\textwidth]{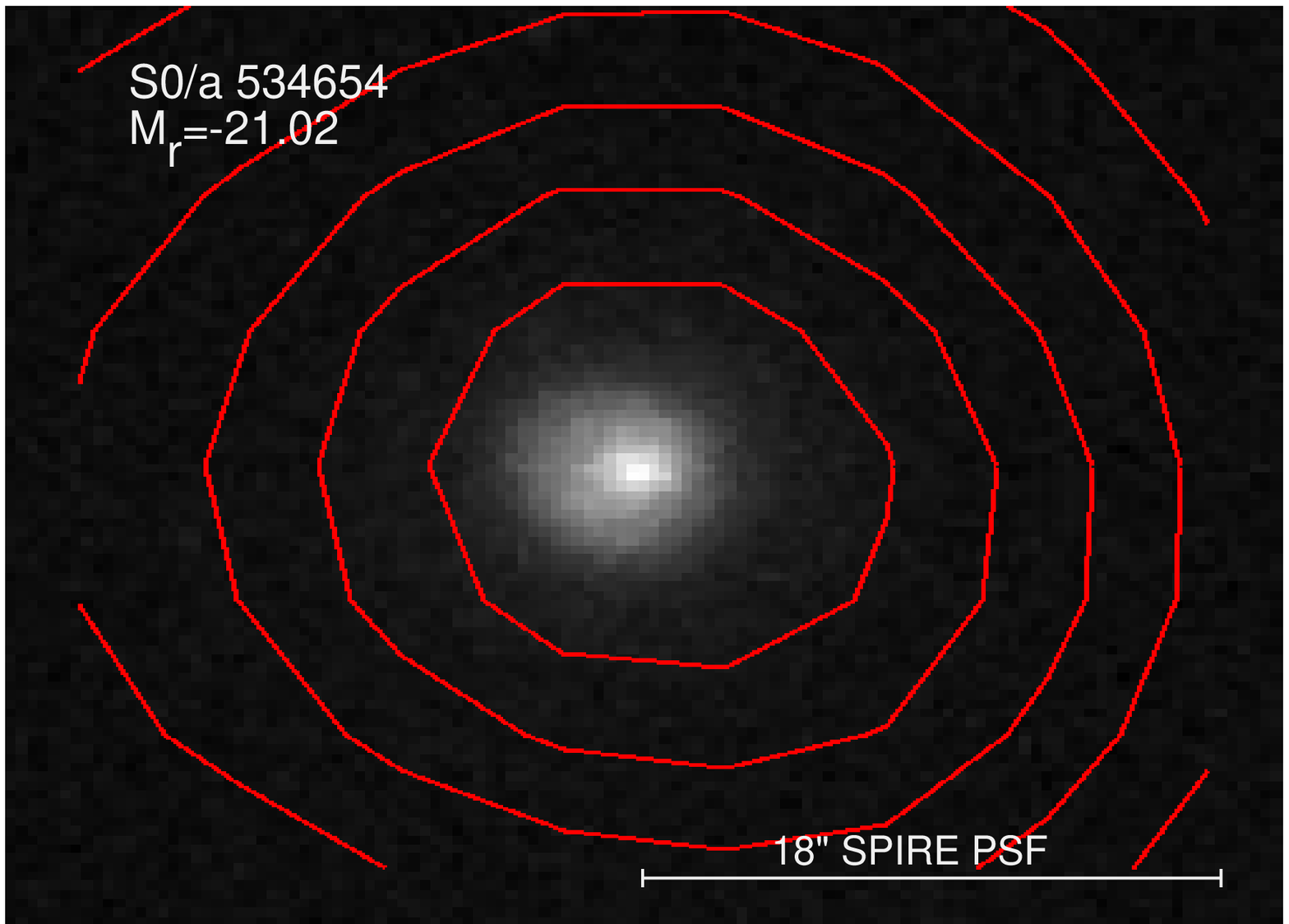} &
       \includegraphics[width=0.3\textwidth,height=0.3\textwidth]{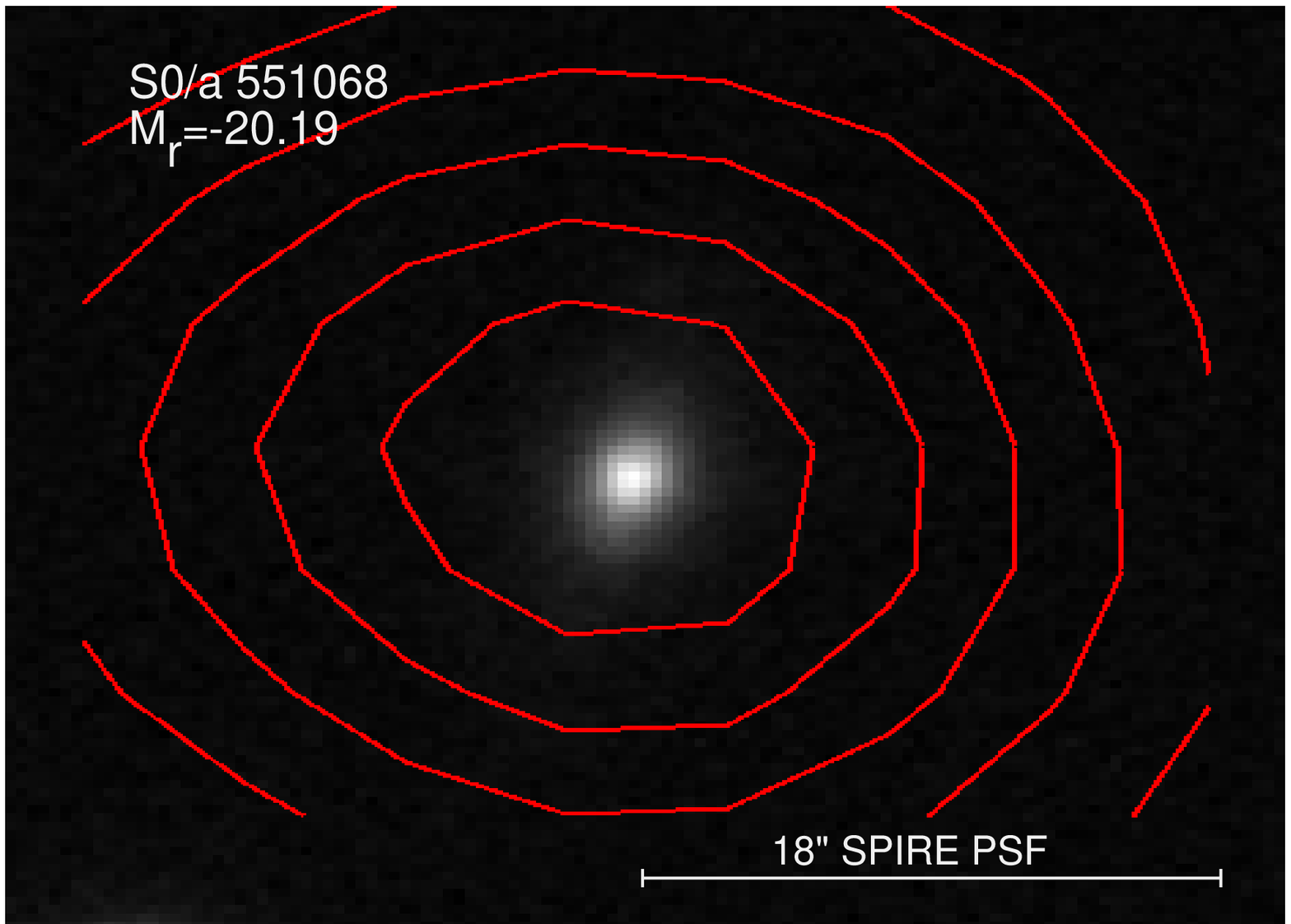} &
       \includegraphics[width=0.3\textwidth,height=0.3\textwidth]{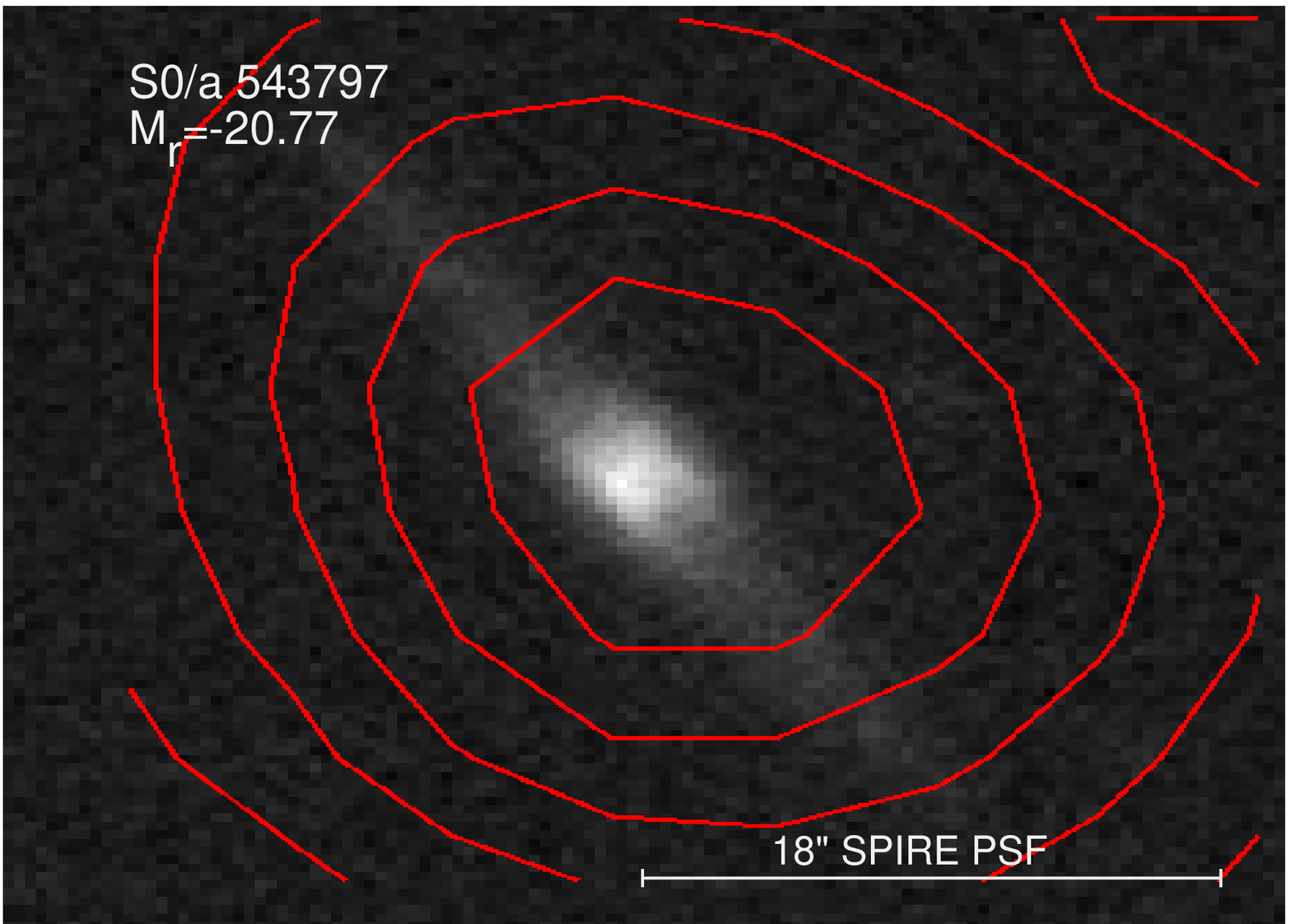} \\       
   \end{array}$
 \end{center}
 \caption{Example images of sub-mm detected H-ATLAS galaxies with E (top) and S0 (bottom) classifications. The images are 40'' by 40'' SDSS $g$-band images with superimposed H-ATLAS 250$\mu$m contours in red. These contour levels represent $\sim$15, 35, 55, 80 and 100$\%$ of the 250$\mu$m flux. Galaxy classification, catalogue ID and absolute $r$-band magnitudes are shown on the top-left of the images. The 18$^{\prime\prime}$ SPIRE FWHM PSF, at 250$\mu$m, is also shown in these images. } \label{fig:GAMApics}
 \end{figure*}

The ETG sample was constructed from the H-ATLAS detected E and S0 (which include both S0, S0a and Sa galaxies) galaxies from within this classified set of galaxies, with additional criteria imposed to remove any potential spiral structure, edge-on disks, or small objects which may be barely-resolved and thereby possibly misclassified. Galaxies with AGN and LINER signatures in the optical BPT diagram \citep{baldwin_1981} were also removed, so as to only consider galaxies with a FIR/sub-mm SED dominated by thermal dust emission. This resultant sample contains 73 Es and 147 S0s, a few examples of which are shown as $g$-band SDSS cutouts in Fig.~\ref{fig:GAMApics}. Please see A13 for details of comparisons of properties between Hubble types, for the H-ATLAS sample.

A13 showed that the H-ATLAS sub-mm detected ETG sample had unusual characteristics in comparison to the undetected ETG. In particular, both optical and UV-optical colours were typically quite blue, indicating some ongoing star formation in these systems (see also \citealp{rowlands_2012}). The galaxy light profiles showed more exponential (or less centrally concentrated) luminosity distributions, which might indicate some recent merging activity, or may be an effect of dust attenuation, or a faint disk. Finally, an investigation of nearest neighbour galaxy surface density revealed that these ETG inhabit sparser environments than the non-detected ETG.

ModBBs with emissivity spectral index $\beta$=2 and 350$\mu$m mass absorption coefficient  $\kappa_{350}$=4.54 cm$^{2}$g$^{-1}$ \citep{dunne_2011} were fit to the PACS (100 and 160$\mu$m) and SPIRE (250, 350 and 500$\mu$m) data for 188 galaxies (see A13). These parameters are fixed to these values throughout this paper. A13 report a range of rest-frame dust temperature between 9-30 K and a range of dust masses between 8.1$\times$10$^{5}$-3.5$\times$10$^{8}$M$_{\odot}$, with a mean dust-to-stellar mass ratio of log$_{10}$(M$_{d}$/M$_{*}$)=-3.37. These results are the key parameters which will be investigated within this paper, in comparison to the nearby ETG in the Virgo Cluster. A summary of the characteristics of H-ATLAS and HeViCS samples (described in $\S$\ref{sec:hevicssample}) is given in Table \ref{tab:parameters}.
       
 	\begin{table}
   \centering
   \hspace*{-0.4in}
      \begin{tabular}{l c c c c}
      \hline
      \textbf{Parameter}	&	\multicolumn{2}{c}{\textbf{H-ATLAS Det.}}	&		\multicolumn{2}{c}{\textbf{HeViCS Det.}}	\\
      & \textit{min} & \textit{max} & \textit{min} & \textit{max}  \\ 
      \hline\hline
      Sample Size (gals)				&	\multicolumn{2}{c}{220}	&	\multicolumn{2}{c}{33}\\
      Distance$_{L}$ (Mpc)					&	57.2	&	265.4		&	17.0	&32.0\\
      log$_{10}$(M$_{*}$) (M$_{\odot}$)		&	8.9	&	11.4		&	8.7 &	11.4\\
      M$_{r}$ (mag)				&	-18.2	&	-23.1		&	 -17.4	&	-23.1\\
      m$_{r}$ (mag)				&	17.7	&	13.3		&	14.1&	8.1\\
      m$_{NUV}$ (mag)				&	22.7	&	16.6		&	18.1&	13.8\\
      F$_{250}$ (Jy)				&	 0.033	&	0.770		&	0.013&	7.992\\
      L$_{250}$ (W Hz$^{-1}$)		&	 1.6$\times$10$^{22}$	&	4.3$\times$10$^{24}$&	4.4$\times$10$^{20}$	&	2.8$\times$10$^{23}$\\ 
           log$_{10}$(M$_{d}$) (M$_{\odot}$)		  &	5.91	&	8.54	&	4.48	&	6.67	\\
                     log$_{10}$(M$_{\textrm{d}}$/M$_{\ast}$) 		  &	-4.44	&	-2.13	&	-6.29	&	-3.07	\\
      $\Sigma_{\textrm{gal}}$ (gals Mpc$^{-2}$)				&	0.001	&	37.08	&	29.19	&	463.10	\\
\hline

   \end{tabular}
   \caption{Parameters indicating the types of ETG found in the sub-mm detected H-ATLAS and HeViCS samples. Note that parameters for the HeViCS sample only include the 33 ETG with M$_{r}\le$-17.4. The parameters shown include sample sizes, ranges of luminosity distances, stellar masses, $r$-band absolute magnitude, $r$ and NUV band apparent magnitudes, total 250\,${\mu}$m fluxes and luminosities. Dust mass, dust-to-stellar mass ratio and environmental density ranges are also shown, calculated as described in the main text.}
   \label{tab:parameters}
\end{table}

\subsection{HeViCS Sample}\label{sec:hevicssample}

       \begin{figure*}
\begin{center}$
\begin{array}{ccc}
      \includegraphics[width=0.3\textwidth,height=0.3\textwidth]{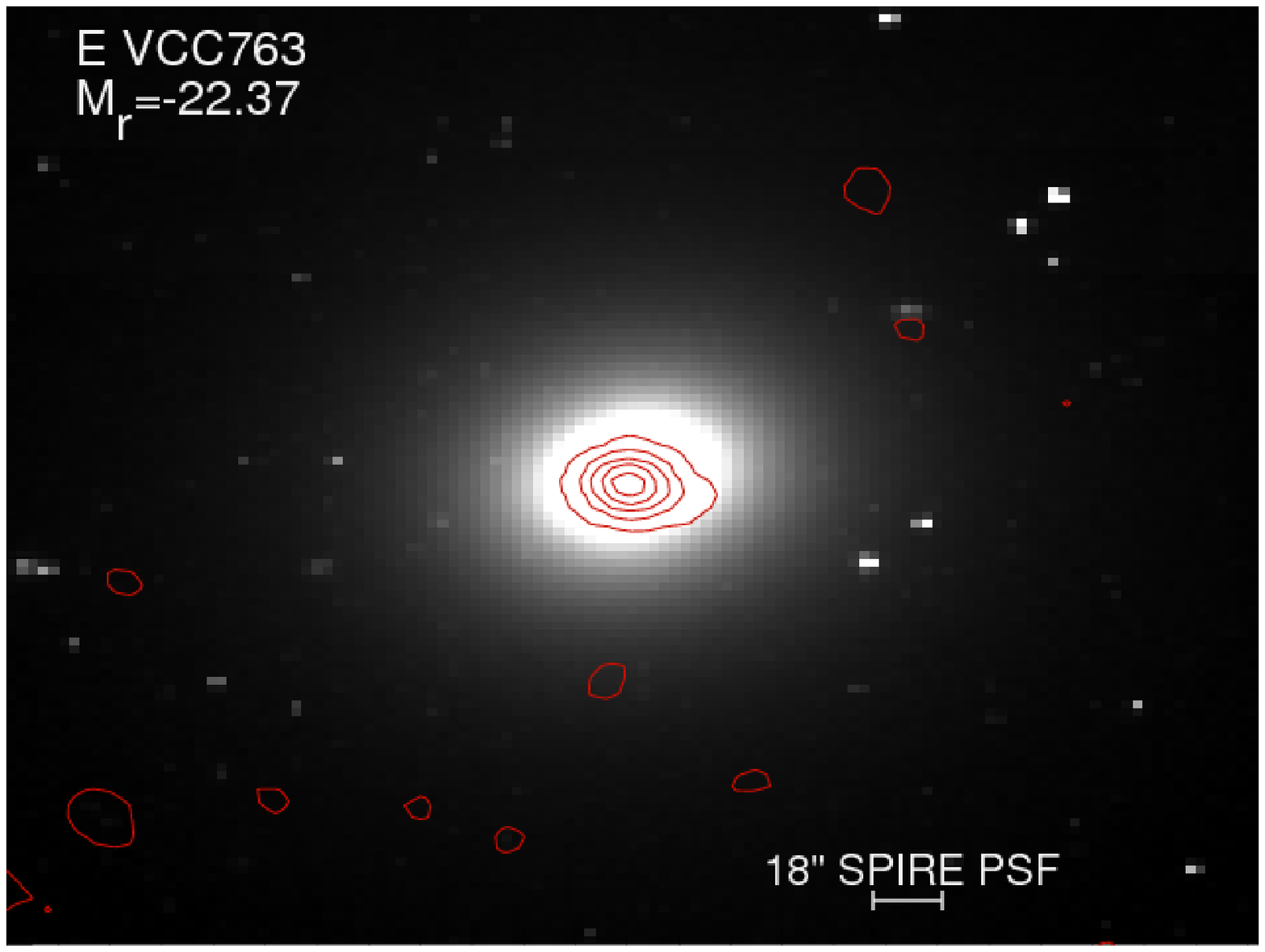} &
   \includegraphics[width=0.3\textwidth,height=0.3\textwidth]{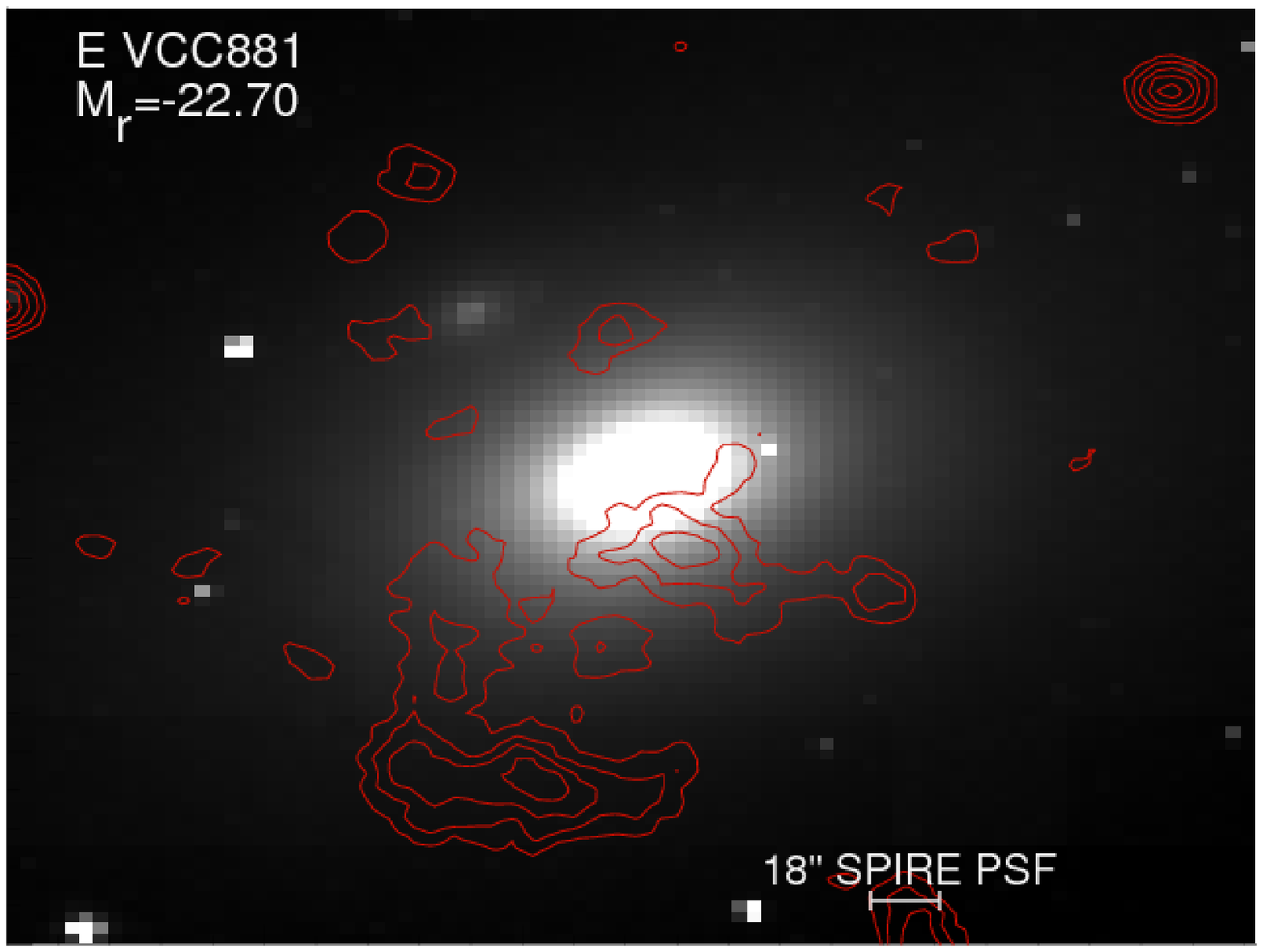} &
    \includegraphics[width=0.3\textwidth,height=0.3\textwidth]{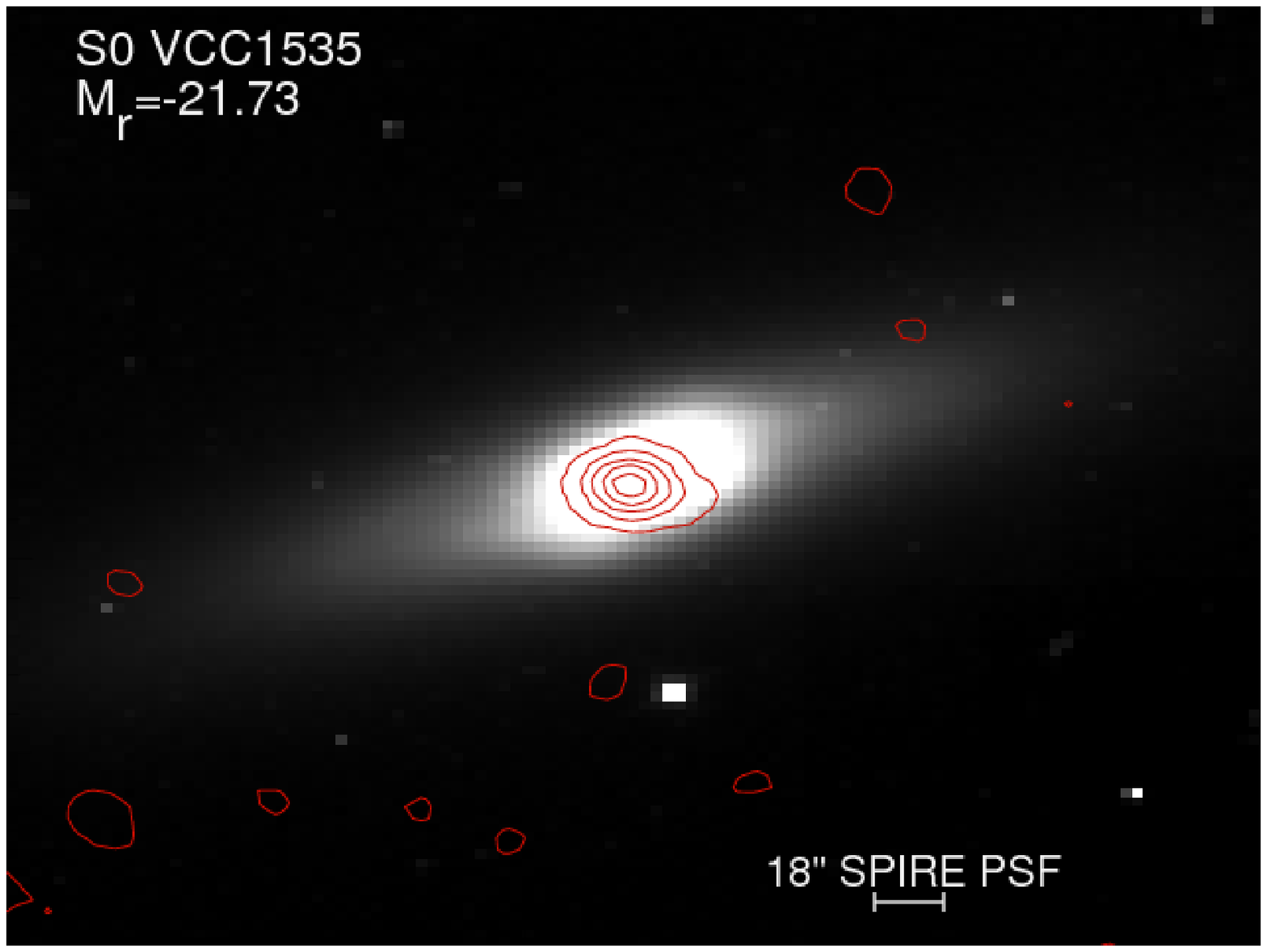} \\
      
   \end{array}$
 \end{center}
 \caption{Example images of sub-mm detected HeViCS galaxies with a variety of classifications. The images are 6' by 6' SDSS $g$-band images with superimposed HeViCS 250$\mu$m contours in red. These contour levels represent the following percentages of the 250$\mu$m flux: 14, 43, 71 and 100$\%$ (VCC763), 18, 36, 55, 73 and 100$\%$ (VCC881), 3, 28, 52, 76 and 100$\%$ (VCC1535). Galaxy GOLDMine classification, identification and absolute $r$-band magnitudes are shown on the top-left of the images. Note that VCC 763 has a synchrotron component. The 18$^{\prime\prime}$ SPIRE FWHM PSF is also shown in these images. }
\label{fig:HeViCSpics}
 \end{figure*}

For HeViCS, S13 utilise an input optical sample from the VCC, constrained by ETG morphology (as compiled in GOLDMine; \citealp{gavazzi_2003}) but not limited in any other respect. Therefore their sample of 910 ETG spans a range of magnitudes, from dwarf systems to the most massive ETG, and contains classifications equal or earlier than S0a-S0/Sa types. 

S13 found 52 ETG by searching for $Herschel$ counterparts within one pixel (6$^{\prime\prime}$) and signal-to-noise (S/N) greater than 5, in the parent sample described above. The reliability of these counterparts is fully discussed in S13. In order to make a fair comparison with the brighter galaxies in the GAMA/H-ATLAS sample, we selected a bright subsample of the 910 ETG of the input HeViCS sample, by applying a cutoff of SDSS M$_{r}\le$-17.4, as with the GAMA galaxies. This subsample is complete for the Virgo Cluster and contains 123 ETG. Out of these, 33 ETG are detected at 250 $\mu$m: their properties are given and contrasted to the H-ATLAS sample in Table \ref{tab:parameters}. Sixteen of the nineteen ETG detected at 250 $\mu$m, which have been removed, have formal GOLDMine dwarf classifications. From here on, these 33 ETG will form the HeViCS detected sample. Correspondingly 90 Virgo ETG with M$_{r}\le$-17.4 are undetected at 250$\mu$m. The magnitude cutoff M$_{r}\le$-17.4 also has the effect of removing all those ETG identified as possible contaminating background sources in S13. Therefore all the 33 HeViCS ETG considered here have secure identifications.  Fig.~\ref{fig:HeViCSpics} shows a few examples of Virgo ETG detected in HeViCS. VCC763 and VCC1535 are shown to have centrally distributed sub-mm emission, but VCC881 is a special case, as the sub-mm contours are quite offset from the galactic centre. This is likely due to the streams linked with this galaxy, indicative of dwarf companion stripping (e.g. \citealp{trinchieri_1991,kenney_2008,janowiecki_2010}).

Apparent $r$-band magnitudes for this ETG sample have been obtained from the work by \citet{cortese_2012b}, where they calculated UV and optical asymptotic magnitudes for the HRS galaxies, some of which are in the Virgo cluster. This provided AB $r$-band magnitudes for 148 HeViCS galaxies; the remaining HeViCS galaxies have magnitudes calculated from the combination of B-band magnitudes from GOLDMine and the average (B-$r$) = 1.02$\pm$0.26 colour obtained from these 148 galaxies. These apparent magnitudes are then converted to absolute magnitudes using the GOLDMine distances and the appropriate galactic absorption.

Stellar masses were estimated for these galaxies using the method of \citet{zibetti_2009} whereby optical (and NIR when available) photometry and synthetic libraries are compared. Dust temperatures and masses were derived from modBB fits to the FIR/sub-mm data. For the 33 massive (i.e. non-dwarf) ETG and the same values for $\beta$ and $\kappa_{350}$ as the H-ATLAS sample, dust temperatures and masses are given as 14.6-30.9 K and 3.0$\times$10$^{4}$-4.7$\times$10$^{6}$M$_{\odot}$. Their mean dust-to-stellar mass ratio is log$_{10}$(M$_{d}$/M$_{*}$)=-3.93. 

\subsection{Sample Comparison}\label{sec:comparison}

There are some clear differences between the two samples which need to be addressed before proceeding with a comparison of their properties. A primary concern is the difference in galaxy distance (see Table \ref{tab:parameters}): HeViCS ETG are located in the nearby Universe at a distance between 17 and 32 Mpc, whereas H-ATLAS ETG are further away within a distance range of 56$\le d_{L} \le$269 Mpc with an average distance of $\sim$195 Mpc. As a consequence, H-ATLAS ETG will have lower spatial resolution, larger luminosity at the optical detection threshold, and a higher dust-mass detection threshold. HeViCS ETG are very well resolved and have lower detection thresholds at all wavebands. For H-ATLAS, this results in the morphological classification not being as detailed as that completed for HeViCS. Therefore H-ATLAS ETG can be identified as either E or S0 galaxies, but we cannot distinguish safely between these two classes, nor detect any dwarf galaxies, which in any case are excluded by the M$_{r}\le$-17.4 limit. Also the completeness at this limit of H-ATLAS ETG is not as good as for the HeViCS ETG for the reasons explained in Section 3.1.

Given that this work will contain a statistical analysis of the properties of the two samples, it is important to consider whether the ETG sample sizes are statistically significant. Additionally, when comparing properties for the samples using statistical tests, it is preferable for the sample sizes to be of similar orders of magnitude in order to obtain a fair analysis. The H-ATLAS sample contains 220 ETG detected in the FIR whereas HeViCS contains 33 ETG (within the H-ATLAS magnitude cutoff of M$_{r}\le$-17.4). Both samples are large enough to run a Kolmogorov-Smirnov (KS) test, which is sensitive to fairly small differences between modest sized samples,
to check whether the populations are similar. The sample sizes themselves are different with the HeViCS sample only containing $\sim$15$\%$ of the H-ATLAS numbers, but the differences are not so large that such a test would be invalid. 

HeViCS dust masses and specific dust masses are lower than those of the H-ATLAS sample by approximately an order of magnitude, even though the morphologies of the galaxies are similar and the stellar mass ranges overlap. This is due to the fact that the closest H-ATLAS ETG (those at z$\sim$0.013 or distance d$_{L}\sim$56 Mpc) are more than three times further away from us than the main Virgo cloud at 17 Mpc, and on average the H-ATLAS ETG are still much further away (see Table \ref{tab:parameters}). Therefore smaller quantities of dust (at least ten times smaller) can be detected in HeViCS ETG than in H-ATLAS. These differences need to be taken into account to avoid possible biased conclusions about the properties of ETG as a class.

Dust appears to be much more concentrated than stars in Virgo ETG, and more luminous ETG have higher dust temperatures (\citealp{smith_HRS},S13). The dust mass does not correlate clearly with stellar mass, while the dust-to-stellar mass ratio anticorrelates with galaxy luminosity. The dusty ETG appear to prefer the densest regions of the Virgo cluster. Contrary to H-ATLAS/GAMA ETG, the HeViCS ETG detected at 250$\mu$m are not bluer than the undetected HeViCS ETG \citep{sperello_2013b}.

The significant difference in 250$\mu$m luminosities between the two samples also needs to be considered, especially given the similarity in optical luminosity between the two samples. We calculate this parameter for all the ETG. H-ATLAS ETG have a range of luminosities from 1.6$\times$10$^{22}$-4.3$\times$10$^{24}$ W Hz$^{-1}$, whereas HeViCS have a range from 4.4$\times$10$^{20}$-2.76$\times$10$^{23}$ W Hz$^{-1}$. We define a threshold luminosity, defined by the H-ATLAS flux limit of 33.5~mJy and maximum distance (z=0.06 $\rightleftharpoons$ 
d$_{L}$=269 Mpc), which equates to 2.77$\times$10$^{23}$ W Hz$^{-1}$. There are 52 ETG (24$\%$) within H-ATLAS with detections below this threshold luminosity; therefore there is some overlap in FIR luminosity between the two samples. It is possible that the difference in luminosities is a direct result of the much larger volume covered by H-ATLAS/GAMA, which gives a larger chance of seeing rarer objects.

\section{Exploring Environments}\label{sec:section4}

Galaxies in the HeViCS sample are, by design, located in the Virgo Cluster. Although the cluster is an overall dense environment, the density is not homogeneous and will vary with position throughout the cluster. Conversely, because the ETG in the H-ATLAS sample were taken from a wide area of sky over a volume of redshift, they most likely belong to a range of environments, thus reaching lower densities. Therefore, a quantitative comparison of the environments inhabited by the ETG in both samples requires some form of estimation of the environmental density performed in a consistent manner.

\subsection{Nearest Neighbour Densities}

To calculate environmental densities we utilise nearest neighbour surface densities. This has already been done to some extent for the H-ATLAS sample (\citealp{brough_2013}, A13), although a bright magnitude limit of M$_{r}\le$-20 was imposed that may not accurately sample the true densities of these ETG. Nearest neighbour densities are now calculated which do not incorporate so bright a magnitude limit.

Chris Beaumont's IDL library\footnote{http://www.ifa.hawaii.edu/users/beaumont/code/} is used to calculate a smoothed map of the coordinates of all the galaxies in the HeViCS and H-ATLAS sample regions respectively, based on the method in \citet{gutermuth_2005}. Although we are only interested in the environments of the sub-mm detected ETG, it is necessary to perform this routine on the entire population of galaxies within these regions to accurately depict the true density; this is the density-defining population (DDP). For every galaxy, the algorithm calculates the distance D$_{N}$ to the N-th closest object and thus the surface density

\begin{equation}\label{eq:surfacedensity}
\Sigma_{gal} = \frac{N}{\pi D_{N}^{2}}.
\end{equation}
	
\noindent The value of N chosen for these calculations is five as this is a good compromise given the effects of survey edges. Additional calculations are then required to convert the units of the surface densities from objects per square degrees to objects per square Mpc. For the HeViCS galaxies, this is a straightforward conversion using the distance of the galaxy.

These calculations are not as simple for the H-ATLAS sample; because of the large redshift range of the galaxies, \citet{brough_2013} and A13 limited the DDP for each sample galaxy to a velocity cylinder $\pm$1000 km~s$^{-1}$ over which the surface density is calculated, so that the latter is not influenced by galaxies at large distances, which clearly cannot have any environmental effect. This is also repeated here. Once each DDP has been created, the procedure described above for the HeViCS densities can be run, and surface densities created.

An additional restriction for these calculations is the imposed magnitude limit on the galaxies used to create the DDP. The Virgo Cluster galaxies can be detected down to much fainter magnitude limits than the higher redshift galaxies, and therefore surface densities for the latter are likely to be underestimated because dwarf galaxies which are detected in the Virgo cluster cannot be detected at higher redshifts. To avoid this, we set a magnitude limit on the DDP of M$_{r}\le$-17.4 for both samples. This is the faintest limit which still ensures completeness also on the H-ATLAS/GAMA sample at $0.013<z<0.06$.

	\begin{figure}
\begin{center}
\hspace*{-0.3in}
 \includegraphics[width=0.6\textwidth]{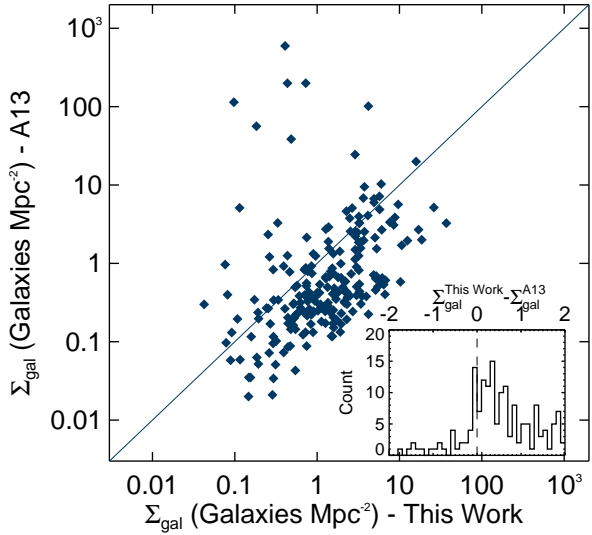} 
 \caption{A comparison of the nearest neighbour densities for the H-ATLAS sample as calculated in this work with those calculated in A13. The key difference in the calculation of these densities is the magnitude limit of the DDPs required to calculate these values. A one-to-one correlation is shown to aid comparison. The inset plot shows the quantitative difference between the two in histogram form.} \label{fig:density}
\end{center}
\end{figure}

We test the calculated surface densities by comparing them with those derived by \citet{brough_2013} and displayed in A13. Fig.~\ref{fig:density} shows a direct comparison between the two parameters, with a mostly linear relation defined. There are some galaxies from  A13 which have higher surface densities than those calculated here: these are upper limits calculated for these galaxies flagged as lying at the edge of the GAMA regions. More importantly, the surface densities calculated in this work are systematically higher than those in A13. This is as expected, as the fainter DDP magnitude limit will include more galaxies in the calculation, resulting in higher densities. 

\subsection{Sample Environments}

There are three key points to be investigated when comparing the environmental densities of ETG in the H-ATLAS sample with those from the HeViCS sample. Firstly, how do the respective sub-mm detected samples differ in environment and is there any overlap? Next this study is extended to all ETG in A13 compared to all ETG in S13. Finally, it is also of interest whether the sub-mm detected versus non-detected ETG in these respective samples vary in environment between themselves and if so, what the sense of this variation is.

\begin{figure*}
\begin{center}

 \includegraphics[width=0.48\textwidth]{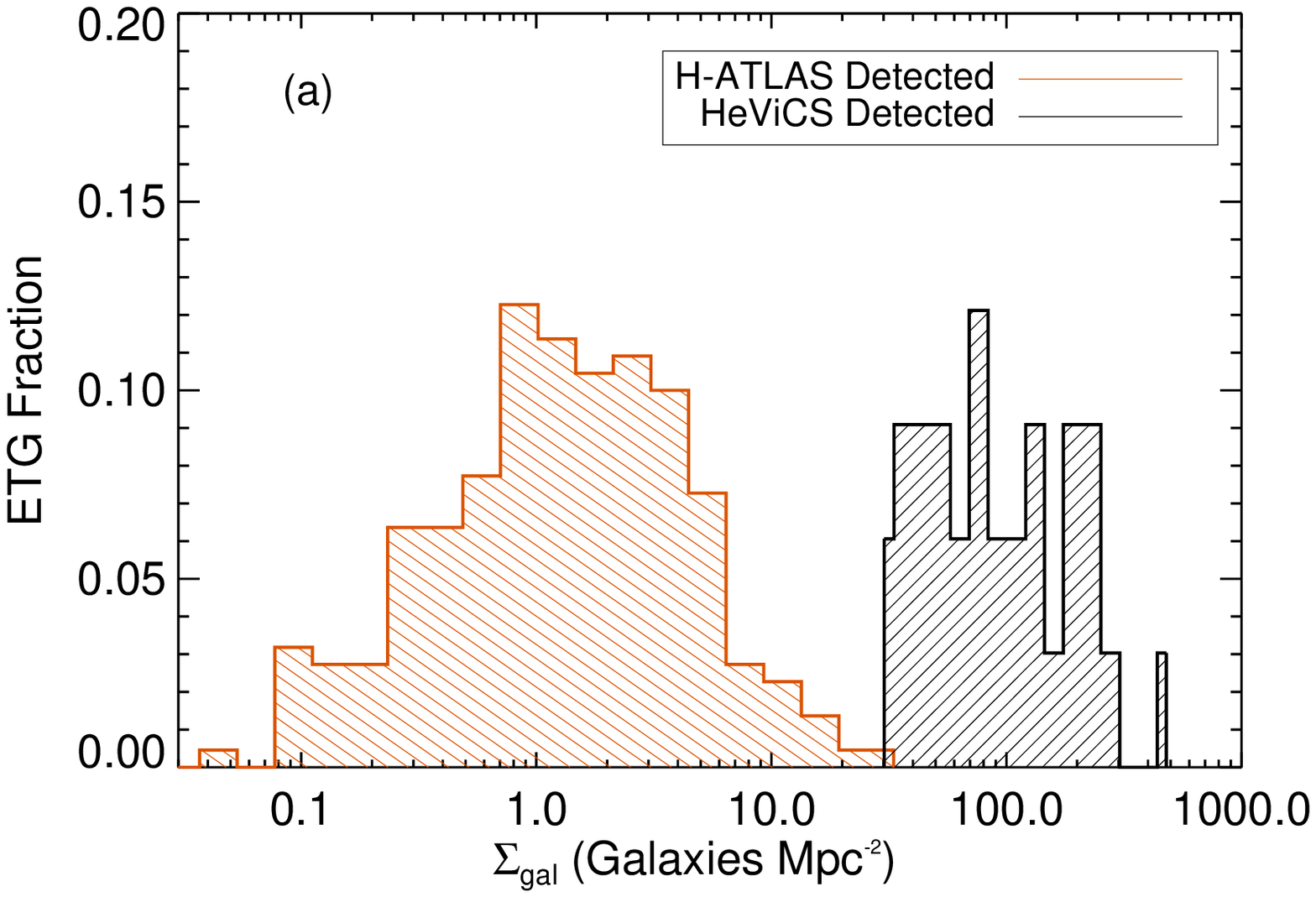}
 \includegraphics[width=0.48\textwidth]{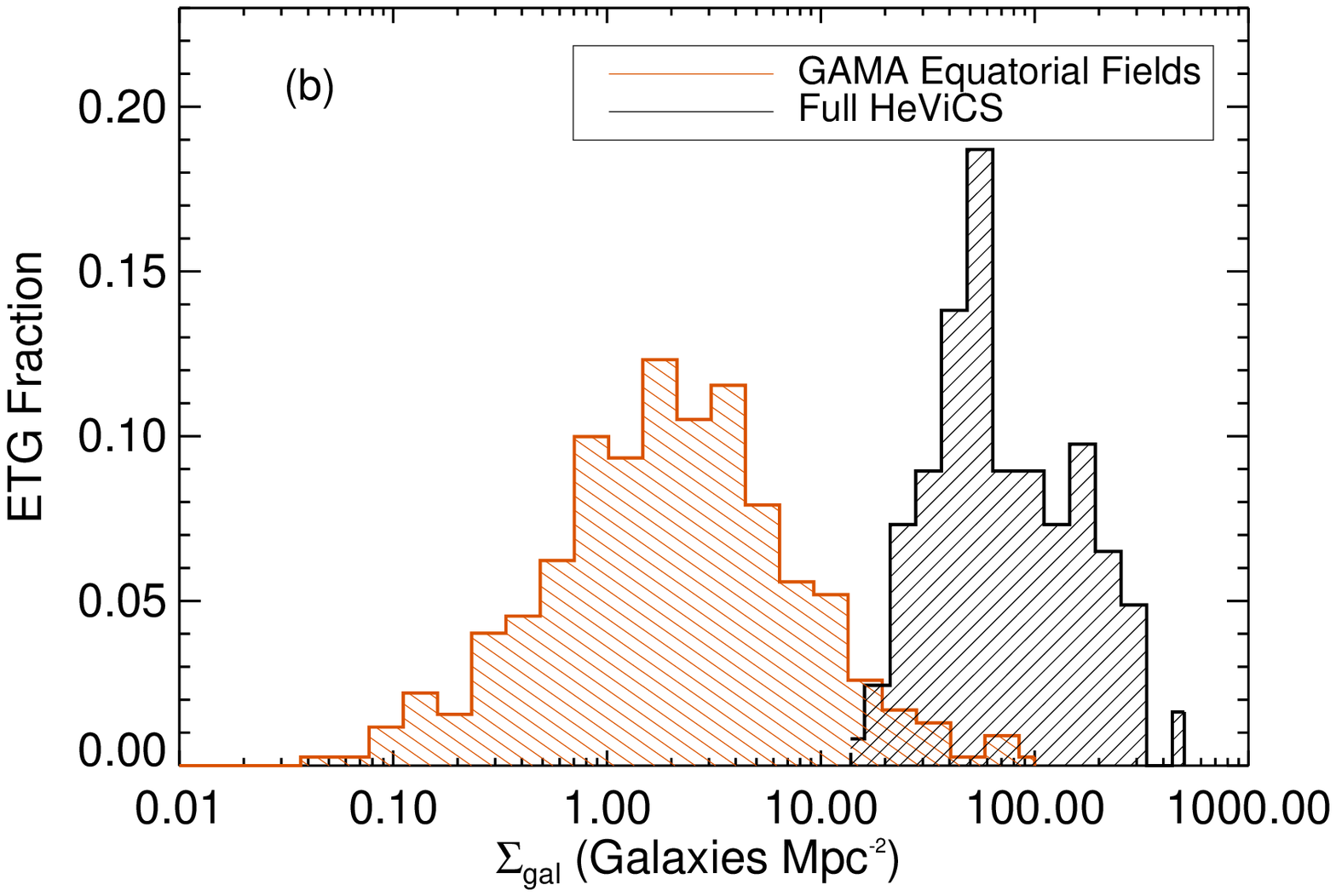}
 \caption{Left panel: normalised distributions of sub-mm detected ETG in H-ATLAS (orange histogram) and HeViCS (black histogram) samples. Right panel: distributions of the combination of sub-mm detected and undetected ETG in H-ATLAS (orange) and HeViCS (black).} \label{fig:densitysamples}
\end{center}
\end{figure*}

\begin{figure*}
\begin{center}

 \includegraphics[width=0.48\textwidth]{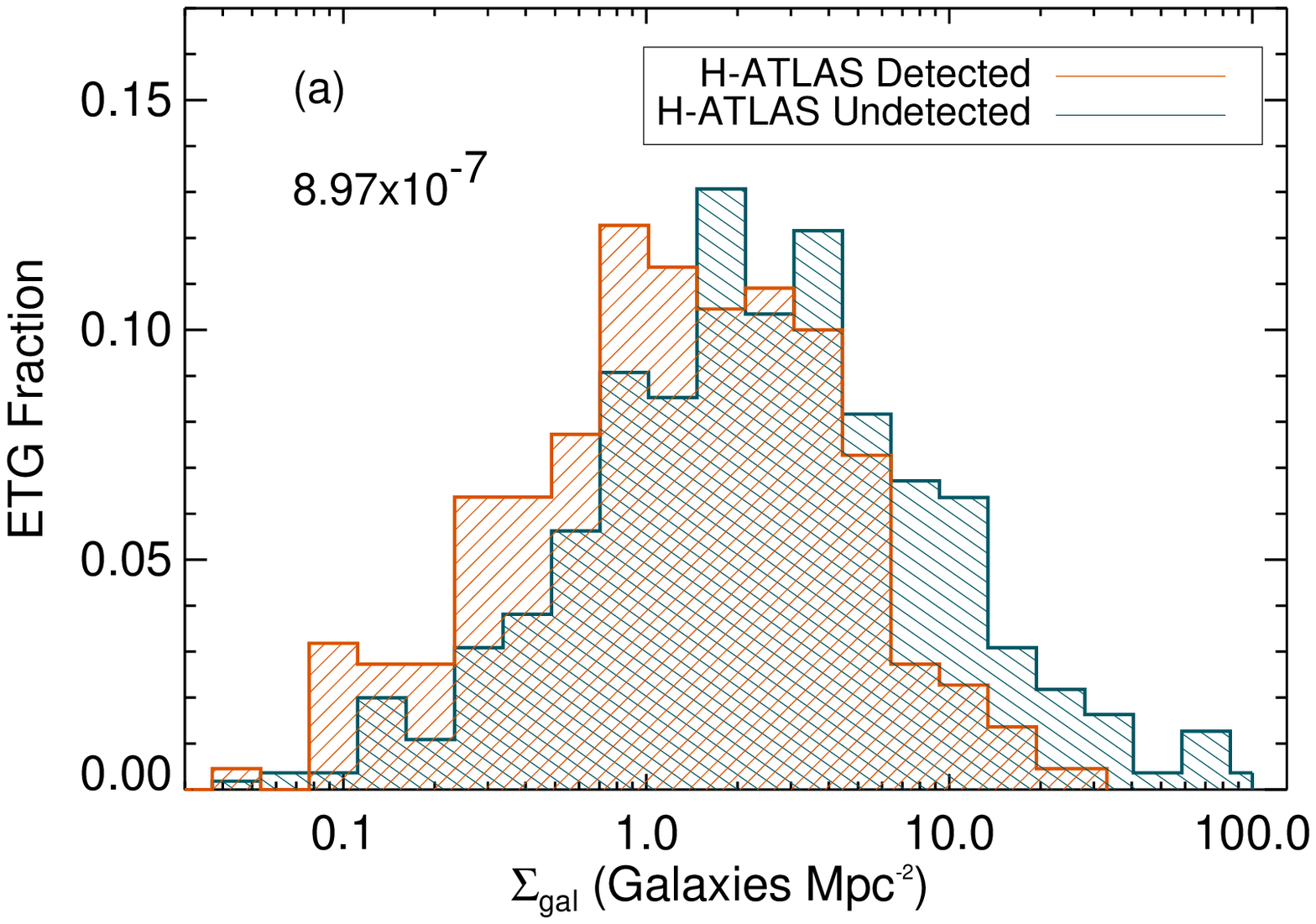}
 \includegraphics[width=0.48\textwidth]{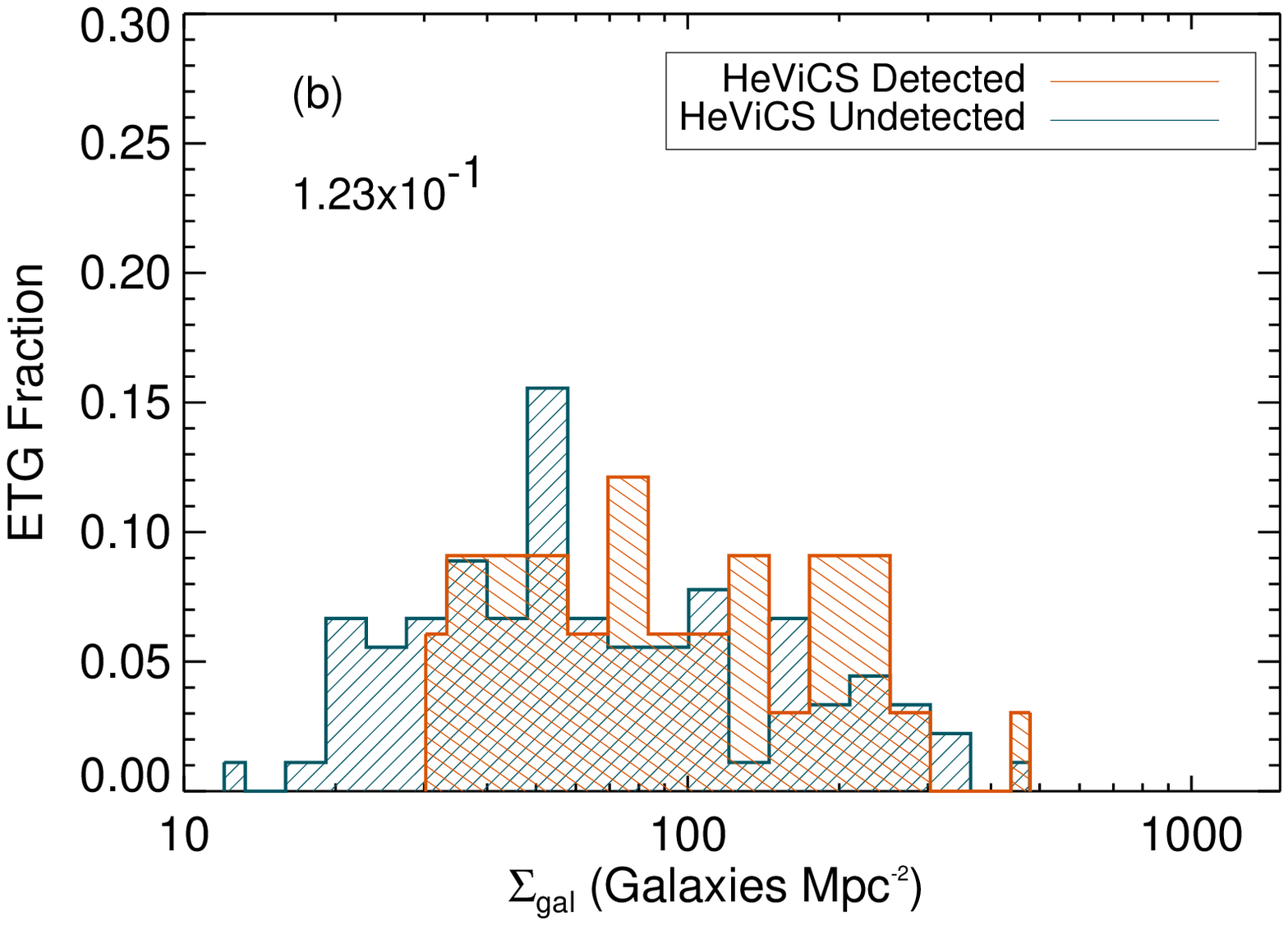}
 \caption{Left panel: normalised distributions of sub-mm detected (orange histogram) and sub-mm undetected (blue histogram) ETG in H-ATLAS. Right panel: distributions of sub-mm detected (orange histogram) and sub-mm undetected (blue histogram) ETG in HeViCS. KS probabilities for whether the presented samples are drawn from the same distribution are shown in the top left of the plots.} \label{fig:betweensamples}
\end{center}
\end{figure*}

Trends of these distributions of densities are investigated in Figs. \ref{fig:densitysamples} and \ref{fig:betweensamples}. A KS test of the sub-mm detected samples' densities in Fig.~\ref{fig:densitysamples}(a) reveals a probability of only 2$\times$10$^{-26}$ of the two distributions being drawn from the same parent distribution. H-ATLAS and HeViCS ETG clearly reside in very different environments, where HeViCS ETG are dominated by the dense cluster environment; the H-ATLAS ETG on the other hand mostly occupy sparse and non-cluster environments.

Examination of the samples including those ETG without sub-mm detections in Fig.~\ref{fig:densitysamples}(b) reveals the probability of ETG residing in the same environments is also zero, yet there is a modest overlap in the environments for the two samples between 20$<\Sigma_{gal}<$100 gals Mpc$^{-2}$. This overlap can mostly be associated with those ETG which are not detected at sub-mm wavelengths and is explored further in Fig.~\ref{fig:betweensamples}. Note that such an overlap is not apparent in Fig.~\ref{fig:densitysamples}(a) for the sub-mm detected samples. The GAMA survey as a whole is deep and wide enough to sample a broad range of galaxy environments, from isolated field galaxies, to pairs, and both small and large groups (e.g. \citealp{robotham_2011}). However, it does not sample well the densest regions of the Universe as found in large clusters, since these are very rare environments. This can be seen in Fig.~\ref{fig:densitysamples}(b), which shows that the GAMA galaxies in the three equatorial fields sampled in A13 do not extend up to the densities found in the Virgo cluster. Thus this ETG study is comparing and contrasting largely different environments.

Fig.~\ref{fig:betweensamples} explores the trend of surface density between sub-mm detected and undetected ETG for the respective samples. KS tests for both sets of distributions indicate that there is a significant difference  between the H-ATLAS distributions ($<1\%$ probability of them being the same), but that the HeViCS distributions are very similar. This indicates no environment density preference within the Virgo ETG (for the subsample of data used here). For H-ATLAS (Fig.~\ref{fig:betweensamples}(a)), the sub-mm detected ETG have lower surface densities with respect to those of the undetected ETG - also shown in fig.~10 of A13. In Fig.~\ref{fig:betweensamples}(b) for the HeViCS sample, the opposite effect is suggested. This latter result was noticed by S13 (and contrasted with the result for HI detections and non-detections), when fainter galaxies were included. The tendency found in HeViCS for dusty ETG to occupy the densest regions is consistent with the higher dust detection rate found for HRS ETG inside the Virgo Cluster than for those outside, particularly for lenticulars \citep{smith_HRS}.

This particular difference may be attributed to the fact that the two samples are environmentally very different (as explicitly shown in Fig.~\ref{fig:densitysamples}), with the H-ATLAS sample occupying sparse environments and the HeViCS sample occupying a high density environment. However, given that both strangulation and ram pressure stripping in dense environments are known to typically remove the ISM from galaxies\footnote{For example, dust stripping has been observed to be ongoing in the Virgo Cluster \citep{cortese_2010a,cortese_2010b}.}, it is expected that sub-mm detected galaxies would be in lower density regions than undetected galaxies. The fact that the Virgo ETG are not exhibiting this behaviour indicates some other processes governing the presence of dust within these systems. Attention could be drawn to the case of M86: a Virgo elliptical which appears to have acquired its ISM via the stripping of gas and dust from a nearby spiral \citep{gomez_dust_2010}. \citet{smith_HRS} also suggested that all their ETG acquired their dust through mergers. Conversely, based on the lack of evidence for externally acquired material, \citet{davis_2011} argue against accretion as a general mechanism for gaining gas and dust in Virgo ETG.

An additional effect which may be contributing to this difference is the ability of HeViCS to detect dust to lower levels than H-ATLAS: S13 amongst others found Virgo ETG with dust masses as low as 10$^{5}$M$_{\odot}$, and these lower dust masses appear to be quite common in the Virgo Cluster. Therefore by definition the dusty Virgo ETG are different to those being found by H-ATLAS. We observe that in the Virgo cluster there are no ETG with a 250$\mu$m luminosity above 2.77$\times$10$^{23}$ W Hz$^{-1}$, which is equivalent to the threshold luminosity of H-ATLAS at the redshift upper limit (z=0.06). This is unexpected given that the samples are matched in optical luminosity (M$_{r}\le$-17.4), however it does explain the differences in dust masses currently being observed. Therefore this difference in environments may be a cause of the differences in dust levels in these ETG (dust is destroyed in denser environments).

Another possible cause for this difference that should be considered is the morphological classification of the ETG. HeViCS ETG have high enough optical resolution that they can be definitively categorised into their separate morphologies. Given that H-ATLAS ETG lie at higher redshifts, their associated classifications cannot be assigned the same level of accuracy as the HeViCS ETG and in fact, \citet{kelvin_2014} group S0 and Sa galaxies together in their classifications. Since specific dust ratio of galaxies systematically increases when moving from early- to late-type galaxies \citep{cortese_2012,smith_HRS}, it is possible that a change in the threshold between ETG and late-type classification can skew the results. It is also well known that earlier-type galaxies prefer denser environments (e.g. \citealp{dressler_1980}). Therefore any spurious LTG which may exist in the H-ATLAS sample are likely to have both high dust-to-stellar mass ratio and sparser environments, thereby skewing the sample in the direction being seen. 

This last possibility can be investigated further by estimating possible contamination levels for the H-ATLAS sample. Original classifications from \citet{kelvin_2014} revealed 999 ETG, of which 285 were only agreed on by two of the three classifiers. Assuming that each classifier has approximately equal weight, then these two-way agreements are estimated to be correct $\sim$2/3rds of the time, leaving $\sim$95 incorrect classifications (LTGs misclassified as ETGs). Additional criteria in A13 for creating the ETG samples included size and flattening (which removed 32 galaxies from the H-ATLAS sample, of which some may be LTG), plus evidence of spiral structure (which removed a further 22 galaxies from the H-ATLAS sample, which are all LTG). This reduces the potential number of contaminants to about 41 to 73 LTG. This is only $\sim$ 5 to 9 $\%$ of our sample of 771 ETG, whereas we detect 29$\%$ of our ETG sample in the submm. Therefore our observations are not consistent with resulting from LTG contamination alone. These values are estimates.
This question of contamination of the H-ATLAS ETG sample will be better addressed in future through the use of deeper, sharper images from the VISTA and VST surveys covering GAMA areas (e.g. \citet{sutherland_2015}).

\section{Multi-wavelength SED Fits} \label{sec:section5}

The photometric data of both the HeViCS and H-ATLAS sample cover a wide wavelength range. Intrinsic information is encoded in the spectral energy distribution (SED) of each galaxy. Certain wavelengths are directly linked to a single component (e.g. NIR emission traces the old stellar population), while others are  ambiguous (e.g. optical light is influenced by both stars and dust). It is therefore useful to treat the multi-wavelength information of a galaxy at the same time using a complete model.

\subsection{MAGPHYS}\label{sec:multiwavelengthfits}

MAGPHYS - Multi-wavelength Analysis of Galaxy Physical Properties \citep{dacunha_2008}
 is a Bayesian fitting code which is able to model the UV to submm SED of galaxies. The program relies on a multi-component galaxy model to predict the flux in each wavelength. Starting from an initial mass function from \citet{chabrier_2003}, stellar components are evolved in time using the stellar population synthesis (SPS) model of \citet{bruzual_2003}. The interaction of starlight with diffuse interstellar and star-forming region dust is calculated from the two-component \citet{charlot_2000} extinction model. One of the key points of MAGPHYS is the physically realistic imposed energy balance between absorbed starlight and dust emission.

Emission from dust grains is modelled using a series of modBB functions and a fixed template for polycyclic aromatic hydrocarbon (PAH) features, as described in \citet{dacunha_2008}. The free parameters in this dust model are the relative contributions of each component to the total IR emission and the temperatures of the warm circumstellar dust $T_\mathrm{W}^\mathrm{BC}$ and cold interstellar dust $T_\mathrm{C}^\mathrm{ISM}$. We adopt an expanded version of MAGPHYS  (da Cunha, priv. comm.) in the sense that the temperature ranges for warm and cold dust are broadened to $30$ K $< T_\mathrm{W}^\mathrm{BC} < 70$ K and $10$ K $< T_\mathrm{C}^\mathrm{ISM} < 30$ K, respectively. This allows for a wider possible range of temperatures found in some systems. It results in longer computation times, but permits better sampling of cold and low star-forming environments. 

A vast library is constructed by randomly drawing parameter sets from the above model and constructing template SEDs with these sets. The expanded version of MAGPHYS that we use here has 50k optical and 75k infrared templates in the library. For the dust an absorption coefficient of  $\kappa_{350}$=4.54 cm$^{2}$g$^{-1}$, with $\beta=2$ is assumed.The observational SED is then modelled by comparing a library of stochastic models (as described in \citet{dacunha_2008}) to the observed data and weighing the output parameters with the corresponding $\chi^2$, constructing probability distribution functions (PDFs). Depending on the data coverage of the SED, some parameters are more accurately constrained than others. In this paper, we limit ourselves to the parameters listed below:

Cold and warm dust are responsible for the large part of the total dust mass in galaxies (M$_C^{ISM}$ and M$_W^{ISM}$ respectively). Warm dust in birth clouds (M$_W^{BC}$) can also contribute. The contributions of hot dust and PAHs are accounted for in a multiplicative factor of $1.1$. Thus the total dust mass is the sum of these components:
\begin{equation}
M_{dust} = 1.1(M_W^{BC}+M_W^{ISM}+M_C^{ISM}).
\end{equation} \\
Cold dust in the diffuse ISM has an equilibrium temperature represented by $T_C^{ISM}$. \\
Accordingly, the equilibrium temperature of warm dust in birth clouds is represented by $T_W^{BC}$. \\
The total amount of infrared light emitted by dust grains is parametrised in the total dust luminosity $L_{dust}$. \\
The total stellar mass  $M_{\ast}$ as derived from the SPS models. \\
The star formation rate (SFR) is an average of the mass of stars formed per year, during the past $100$ Myr. The underlying star formation law is an exponentially declining SFR starting from the birth of the galaxy. Throughout the lifetime of the galaxy, there is a random chance of starbursts taking place. \\
The ratio of the SFR and $M_{\ast}$ is then called the specific star formation rate (sSFR). \\
The time at which the galaxy is formed, $T_\mathrm{form}$, is defined as the age of the oldest stellar population. \\
The time at which the last starburst ended, $T_\mathrm{lastb}$.

     \begin{figure*}
     \centering
        \includegraphics[width=\textwidth]{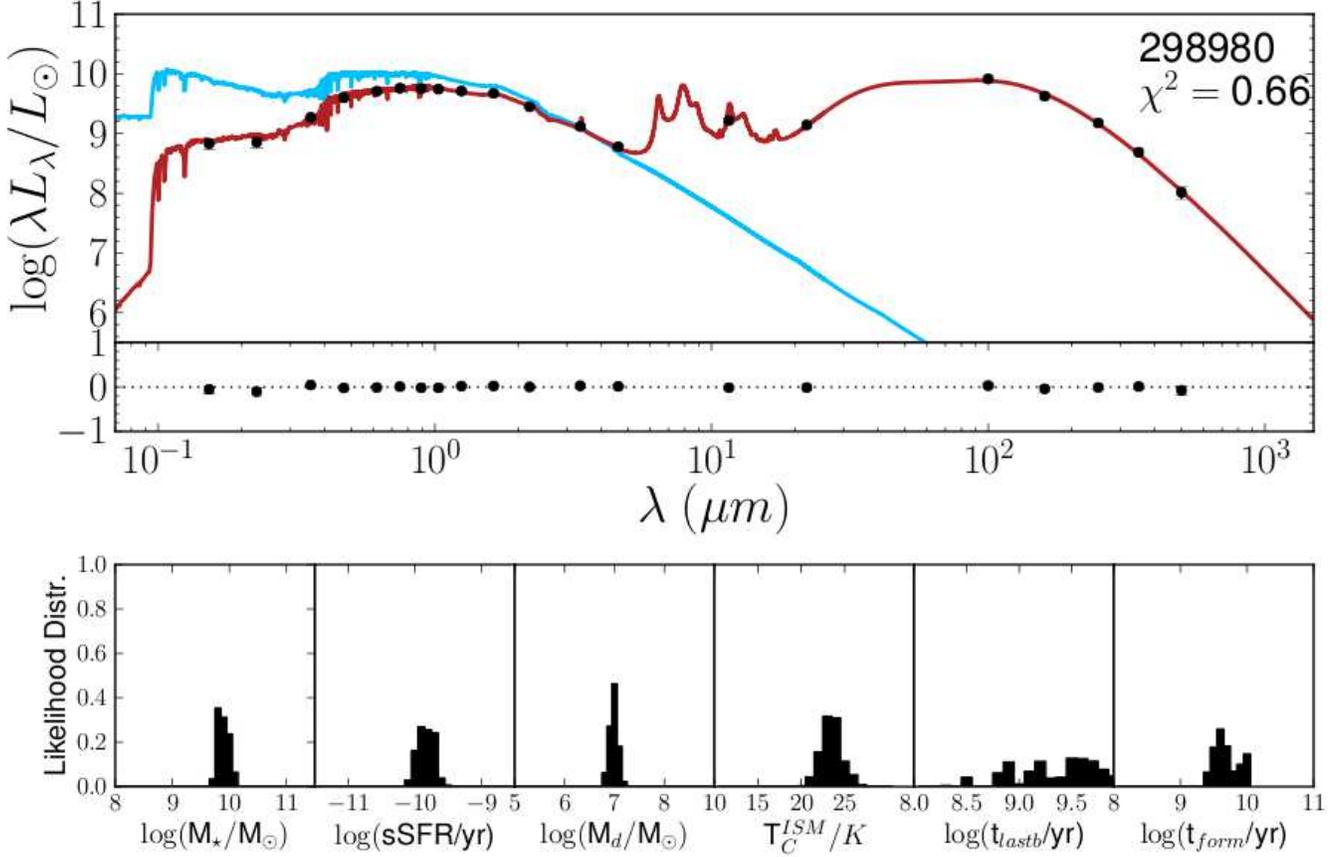} 
    \caption[MAGPHYS Fit for elliptical galaxy 298980]{ MAGPHYS rest-frame SED fit to H-ATLAS elliptical galaxy 298980. The SED is fit to observed photometry in the FUV, NUV, SDSS $ugriz$, UKIDSS YJHK, WISE W1, W2, W3 and W4, PACS 100, 160$\mu$m and SPIRE 250, 350 and 500$\mu$m wavebands (black points). The red line shows the overall attenuated model fit, whilst the blue line shows the unattenuated optical model. Below are the residuals from the fitted points, and below these are the likelihood probability functions for parameters of this elliptical galaxy.}
    \label{fig:sed1}
 \end{figure*}

\subsection{Data coverage of the SED}

Good coverage of the panchromatic SED is desirable to ensure reliable modelling. The GAMA/H-ATLAS dataset comprises of self-consistent photometry based on a standard format. Therefore the H-ATLAS ETG sample has the following data coverage of the SED: GALEX FUV and NUV, SDSS $ugriz$, UKIDSS YJHK, WISE W1-W4, PACS 100 and 160$\mu$m and SPIRE 250, 350 and 500$\mu$m. Note that some galaxies are missing data as detailed below:

\begin{description}
\itemsep0em 
\item[] GALEX FUV: 38 galaxies (17$\%$)
\item[] GALEX NUV: 36 galaxies (16$\%$)
\item[] WISE W1-W3: 5 galaxies (3$\%$)
\item[] WISE W4: 90 galaxies (41$\%$)
\item[] PACS 100$\mu$m: 19 galaxies (9$\%$)
\item[] PACS 160$\mu$m: 26 galaxies (12$\%$)
\end{description}

\noindent Missing galaxies in these wavebands are due to an inability to match the optical source to a counterpart in the specific waveband (i.e. no detection). The 22$\mu$m WISE W4 band in particular suffers from a low detection rate due to the waveband's low signal-to-noise. Data in other wavebands is complete for all the galaxies.

     \begin{figure*}
     \centering
        \includegraphics[width=\textwidth]{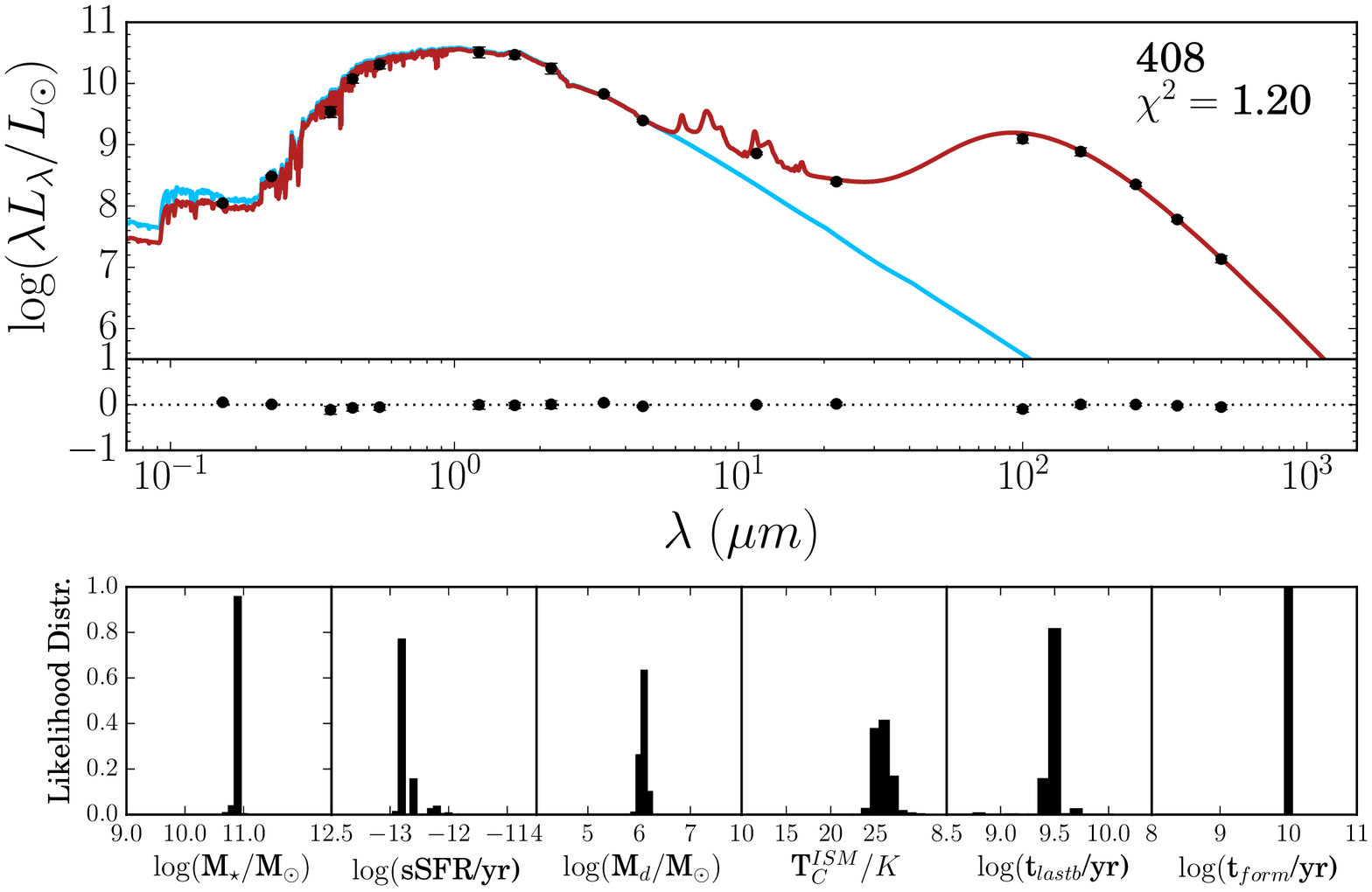} 
    \caption{ MAGPHYS rest-frame SED fit to HeViCS elliptical galaxy VCC 408. The SED is fit to observed photometry in the FUV, NUV, UBV, 2MASS YJHK, WISE W1, W2, W3 and W4, PACS 100, 160$\mu$m and SPIRE 250, 350 and 500$\mu$m wavebands (black points). The red line shows the overall attenuated model fit, whilst the blue line shows the unattenuated optical model. Below are the residuals from the fitted points, and below these are the likelihood probability functions for parameters of this elliptical galaxy.}
    \label{fig:sed2}
 \end{figure*}

In the case of the HeViCS ETG, the UV part of the spectrum is covered by fluxes from the GALEX catalogue except for the 12 galaxies which are also in HRS. For those, fluxes from \citet{cortese_2012} were used. 
Optical and NIR filters include UBV \& JHK from the GOLDMINE archive \citep{gavazzi_2003}. 

WISE \citep{WISE} detected almost all galaxies in the HeViCS sample during its all-sky survey. Unfortunately archival fluxes caused inconsistencies in the SED fits. The MIR is an ambiguous regime, with both emission from old stars and hot dust. Sufficient sampling and correct flux determination are vital to disentangle both components in an SED fit. We therefore chose to redo the flux measurements on the archival WISE images and minimise the contamination of foreground stars.
Appendix \ref{app:wise} gives more detail on the measurements.  

Herschel observations at $100 - 500 \; \mu$m were taken from S13 and complete our dataset.

\subsection{SED Results}

MAGPHYS was used to fit energy balance models to each of the HeViCS and H-ATLAS ETG, as described above. Figs. \ref{fig:sed1} and \ref{fig:sed2} show example MAGPHYS fits to one of the H-ATLAS and HeViCS ETG respectively, with the resultant PDFs for a variety of fit parameters shown in the lower panels. MAGPHYS cannot include a synchrotron component and is therefore unable to correctly fit the four dusty HeViCS ETG with such a component, and therefore these are excluded from the multi-wavelength analysis.

In order to gain some insight on the goodness-of-fit for each galaxy, the 29 (33 minus 4 synchrotron galaxies) HeViCS fits were visually inspected and assigned a flag for `good' or `poor' fit. Four galaxies were assigned `poor fit' status - each of these fits had an associated reduced $\chi^{2}$ value\footnote{The MAGPHYS $\chi^{2}$ is a constraint on the best fitting theoretical template SED and hence the most likely fit.} greater then 4. This was then chosen as the criterion to assess whether the H-ATLAS fits were `good' or `poor'. Eleven H-ATLAS systems were found to have `poor fits'. These fifteen galaxies from the two samples are also highlighted in future plots to separate them from the rest of the sample. Additionally, all ETG with `poor fits' or a synchrotron radio component are excluded in any further statistical analysis in this section.

\subsubsection{Contrasting Derived Parameters}

\begin{figure*}
\begin{center}
\hspace{-0.4in}
 \includegraphics[width=0.54\textwidth]{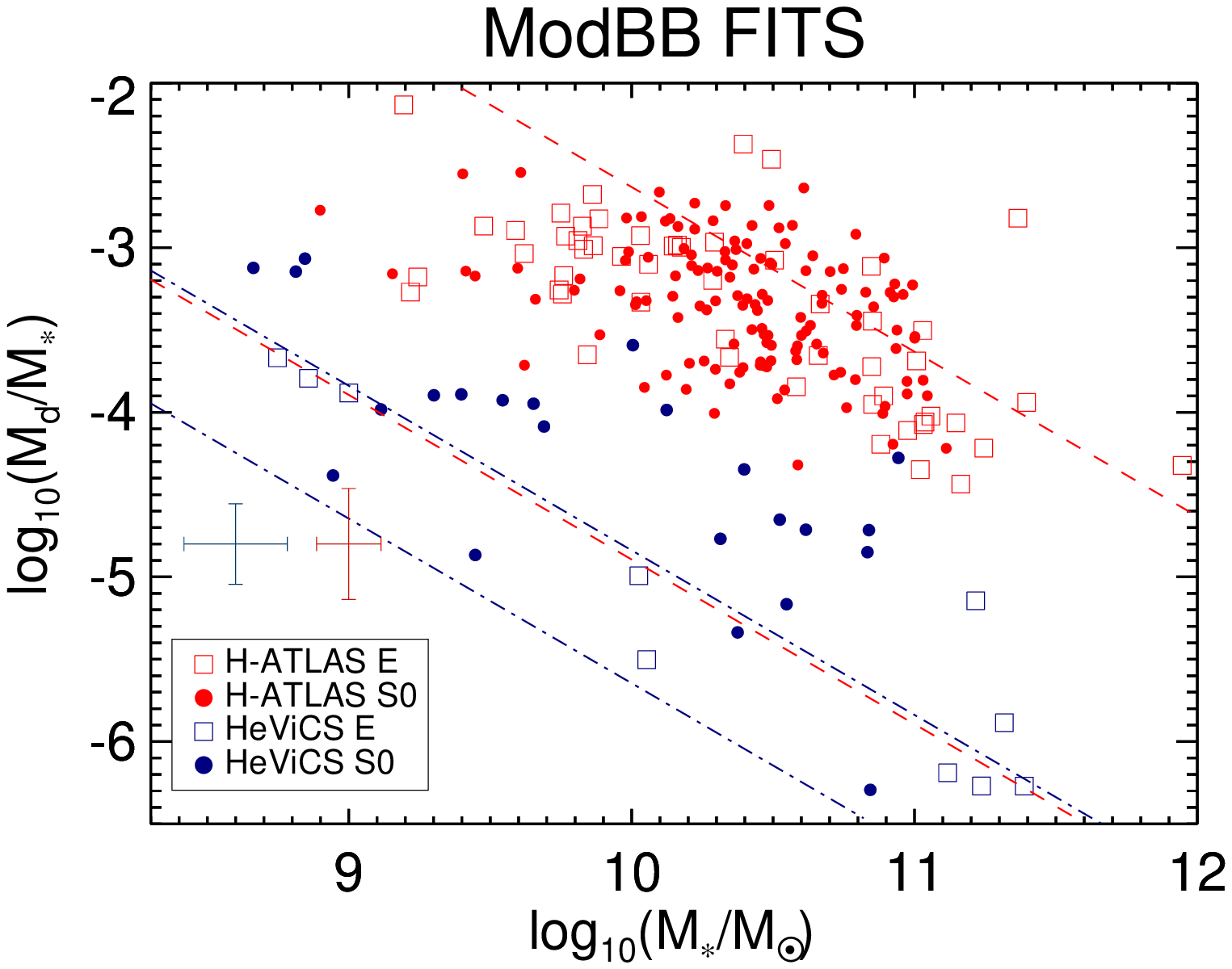} 
\hspace{-0.4in}
  \includegraphics[width=0.54\textwidth]{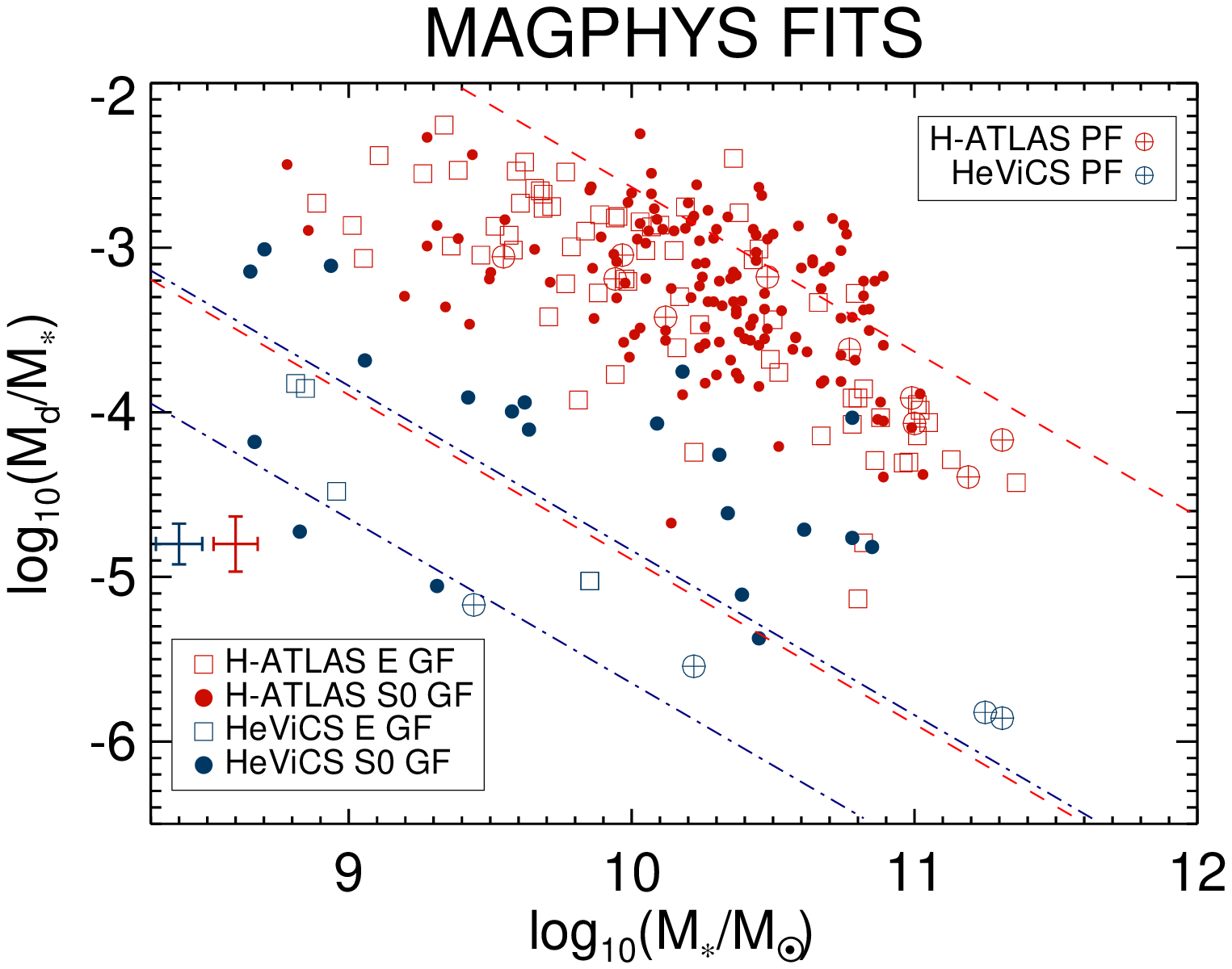} 
 \caption{Dust-to-stellar mass ratio plotted as a function of stellar mass calculated using ModBBs (left panel) and MAGPHYS (right panel). HeViCS (blue points) and H-ATLAS (red points) samples are shown in both plots, and galaxies are subdivided into E (red squares for H-ATLAS and blue squares for HeViCS) and S0 (red dots for H-ATLAS and blue dots for HeViCS) classifications. In the MAGPHYS plot, galaxies with poor fits (PF) are encircled crosses in the samples' respective colours, while all other ponts are good fits (GF). Error bars in the left panel give the mean overall uncertainty on the points from ModBB fits, in the same colours as their respective samples. 
In the left plot diagonal lines represent the range of dust mass detection limits for the samples in their respective colours (see Section 5.3.1 for details). These same lines are included as a guide only in the right plot.
Error bars in the right panel give 1$\sigma$ to each side of the PDF.}
  \label{fig:dustmass}
\end{center}
\end{figure*}

As described in $\S$\ref{sec:SubS} and $\S$\ref{sec:hevicssample}, A13 and S13 fit single modBBs to their FIR/sub-mm data to obtain dust masses and temperatures for their ETG. These dust masses are used to evaluate the dust-to-stellar mass ratio which is plotted as a function of stellar mass in the left panel of Fig.~\ref{fig:dustmass}. The stellar masses are from A13 and S13.
The effects of detection limits are illustrated. The red dashed lines in Fig.~\ref{fig:dustmass} represent the range of dust mass detection limits at the 250 micron flux limit of 33.5 mJy for the H-ATLAS sample, with the lower line for nearby (z=0.013), high temperature (30K) detection limits (as shown in A13) and the upper line for far (z=0.06), low temperature (15K) detection limits. This includes the typical temperature range of most H-ATLAS detections. Below the lower line no detections are expected and above the upper line all sources in this temperature and redshift range are detectable. Between the two lines detectability depends on the distance and dust temperature of sources, with decreasing likelihood of detection going towards the lower line. The blue dot-dashed lines represent the corresponding limits for the HeViCS 250 micron flux limit of 25.4 mJy, the distance range of HeViCS sources (17 to 32 Mpc), and considering the effects of different dust temperatures for the various morphological types from section 4.1 of S13. The specific dust mass detection limits are lower for the HeViCS sample, which is nearby, mostly at a distance of 17 Mpc.
Fig.~\ref{fig:dustmass} shows a key difference in the normalised dust levels of the two ETG samples. It indicates that the HeViCS ETG have less dust, by a factor of 10 or more, than the H-ATLAS ETG.  This difference is due to the much lower dust detection limit of HeViCS, whose galaxies are approximately ten times closer than the H-ATLAS galaxies. However, it remains to be understood why there are no HeViCS ETG as dusty as the dusty H-ATLAS ETG, at fixed stellar mass.

  A two sample KS test can be applied to the left plot in Fig.~\ref{fig:dustmass} by considering a diagonal line parallel to the limits and above which there are no HeViCS detections. This line goes through VCC1535 (at log$_10$(M$_{\ast}$)=10.94 and log$_{10}$(M$_{d}$/M$_{\ast}$)=-4.28 in Fig.~\ref{fig:dustmass} left plot). 162 H-ATLAS/GAMA ETG lie above this line, out of 771 GAMA ETG with M$_r<-17.4$mag. For the Virgo sample there are 123 (including all detected and undetected) ETG with M$_r<-17.4$mag in the HeViCS area, which all lie somewhere below the diagonal line. Therefore the KS statistic is 0.21, which for these optical sample sizes has a probability $<0.1\%$ of them being the same. Thus the two samples differ significantly from each other in Fig.~\ref{fig:dustmass}. More studies of ETG in other clusters would help to verify if this is a difference due to environmental density.

\begin{figure}$
\begin{array}{l}
\vspace{-0.1in}
     \vspace{0.2in}
    \includegraphics[width=0.5\textwidth]{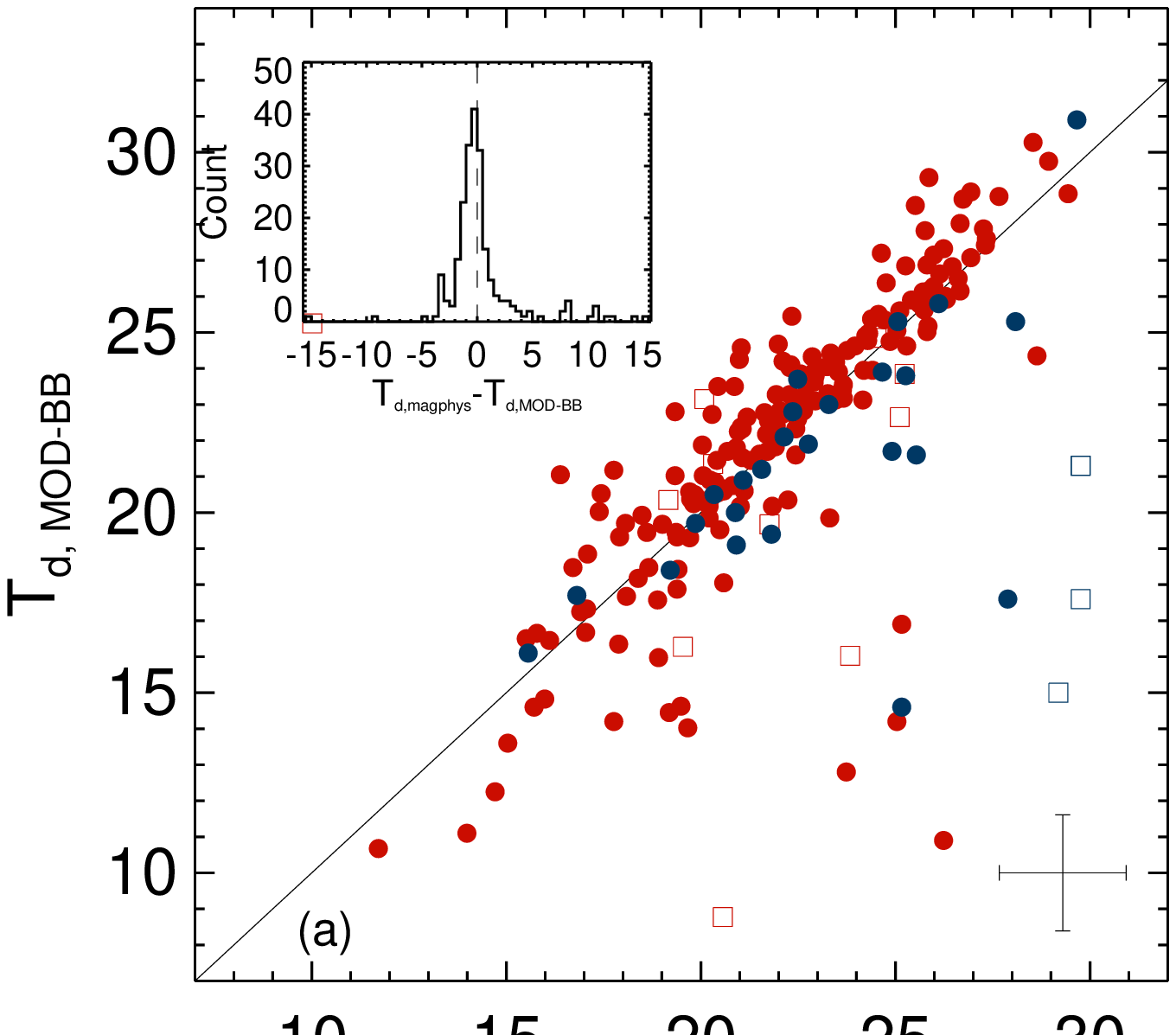}\\
     \vspace{0.2in}
     \includegraphics[width=0.5\textwidth]{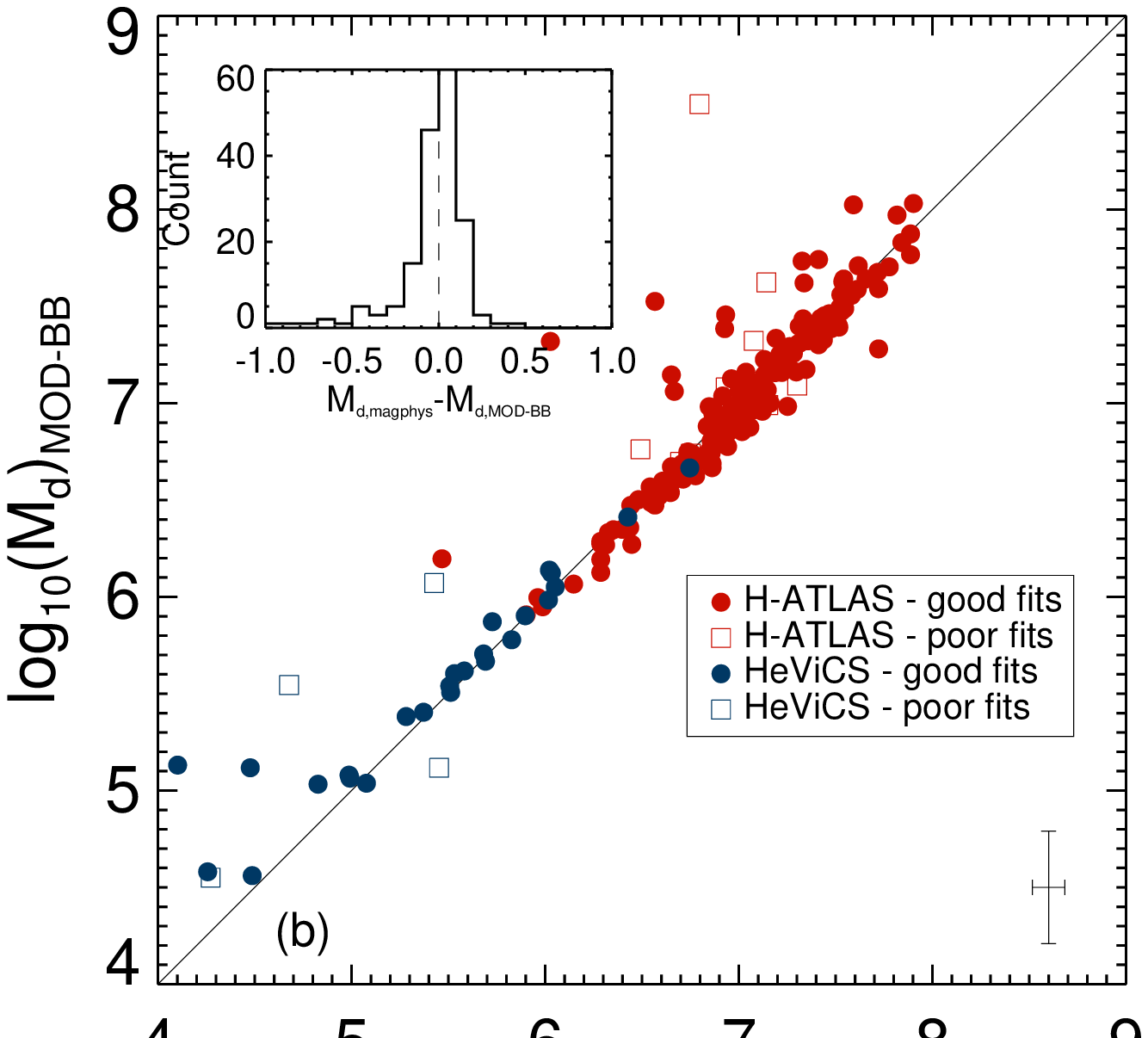} \\
     \vspace{0.1in}
     \includegraphics[width=0.5\textwidth]{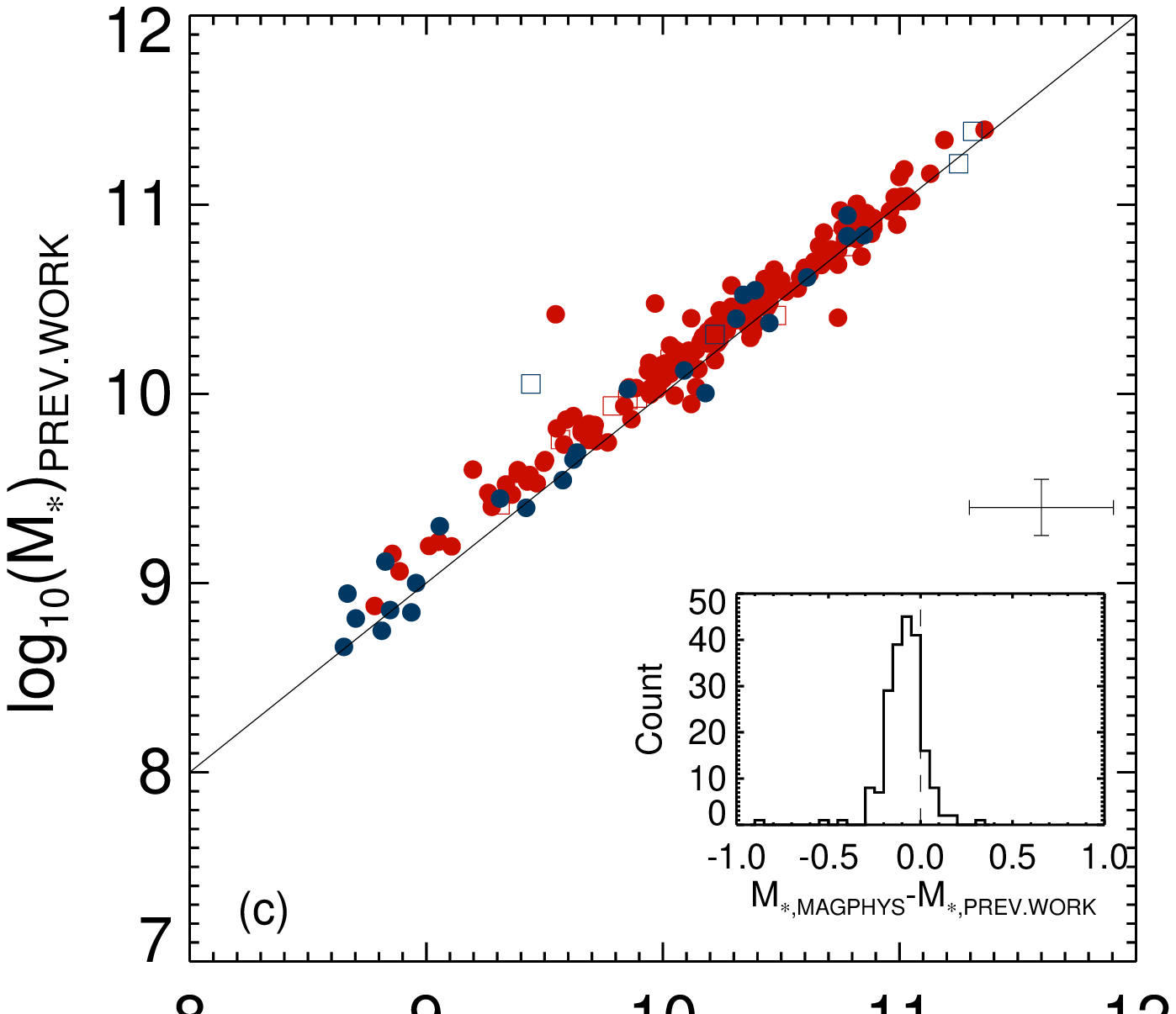}

     \vspace{0.1in}
     \end{array}$
    \caption{ Comparison of derived cold dust temperature (top), dust mass (center) and stellar mass (bottom), where the y-axis represents values from modBB fitting and the x-axis represents MAGPHYS fitting results. For the stellar mass comparison, the y-axis represents stellar mass values from A13 and S13. The H-ATLAS sample is plotted in red and HeViCS ETG are plotted in blue. Galaxies with poor fits ($\chi^{2}\ge$4.0) are shown as open squares. The solid line is a one-to-one line and error bars represent the mean 1$\sigma$ errors in either direction for both samples. Inset plots show the difference offset from zero for each parameter.}
    \label{fig:comparedusttemp}
 \end{figure}
 
Normalised dust mass calculated from MAGPHYS is shown plotted against stellar mass in the right panel of Fig.~\ref{fig:dustmass}, where the stellar masses are also from MAGPHYS. Both ModBB and MAGPHYS plots are shown side-by-side for ease of comparison. Qualitatively there is little difference between these distributions. 
There is a negative trend found for this relationship for both these samples, partly driven by selection effects in the lower left corner. 

Dust properties from MAGPHYS fits are examined and contrasted to the modBB solutions in Figs. \ref{fig:comparedusttemp}(a) and \ref{fig:comparedusttemp}(b). The MAGPHYS most likely\footnote{The most likely value of a parameter is chosen as the median of the probability distribution functions output by MAGPHYS.} cold dust temperature and overall dust mass is shown on the x-axis, and modBB solutions are shown on the y-axis for both samples of ETG. As previously stated, poor fits are shown as different symbols to separate them from the good fits.
 
 Fig.~\ref{fig:comparedusttemp}(a) shows that dust temperatures are similar, except for a few outliers that vary more significantly between the fitting methods. These tend to be for galaxies without PACS detections. The scatter observed here is in the sense that MAGPHYS assigns higher temperatures to the likely cold dust grain distributions. This difference may result from the fact that MAGPHYS fits multiple temperature components, unlike the single component ModBB. More MIR data coverage would help to constrain the dust temperatures in these outliers.
However these differences in temperatures do not necessarily cause similar scatter in other derived parameters.

 MAGPHYS dust mass shows a better correspondence with dust masses derived from modBB fitting (see Fig.~\ref{fig:comparedusttemp}(b)). There is a slight offset for some galaxies in both samples from the x=y plane - typically the modBB fitting appears to give higher dust masses. This is because MAGPHYS assumes a higher on average dust temperature than modBB fitting, thereby resulting in lower dust masses.

These tests reveal that the dust mass parameter is well described by both a modBB and MAGPHYS, as the mean results for the two samples do not change substantially, nor do the KS-test probabilities differ between the two methods. However, a small difference is found between the stellar masses derived by GAMA/HeViCS teams and those derived by MAGPHYS. Fig.~9(c) shows that stellar masses obtained with MAGPHYS are on average slightly smaller (by $\sim$ 0.1 dex) than those obtained previously, particularly at small masses. \citet{taylor_2011} showed that inclusion of UKIDSS fluxes slightly biases stellar mass measurements in this way. This is a small effect for the current applications. No systematic difference between HeViCS and H-ATLAS ETG is apparent in Fig.~9(c), which demonstrates that the previous stellar masses were also obtained consistently for the two samples.

\begin{figure}
\begin{center}
\hspace{-0.5in}
 \includegraphics[width=0.54\textwidth]{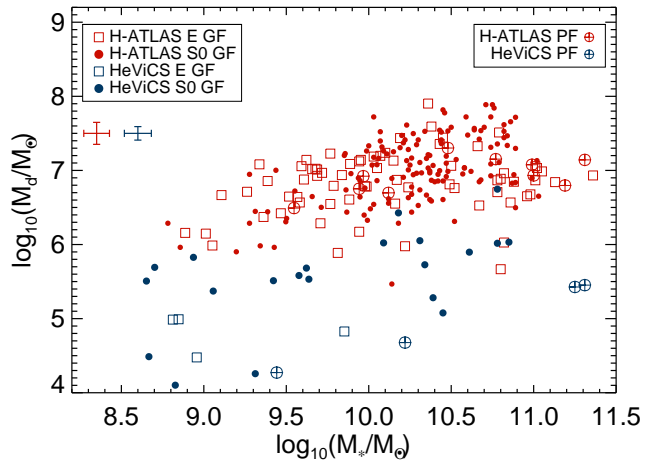}
 \caption[MAGPHYS Dust Mass versus Stellar Mass]{MAGHPYS dust mass plotted as a function of stellar mass. HeViCS (blue points) and H-ATLAS (red points) samples are shown, where galaxies are subdivided into E (red squares for H-ATLAS and blue squares for HeViCS) and S0 (red dots for H-ATLAS and blue dots for HeViCS) classifications. Galaxies with poor fits are encircled crosses in the samples' respective colours.
Error bars give 1$\sigma$ to each side of the PDF, in the same colours as their respective samples. } \label{fig:newdustmassplot}
\end{center}
\end{figure}

We compare the relationship between derived dust and stellar mass for the two samples in Fig.~\ref{fig:newdustmassplot}. Trends are found for both H-ATLAS and HeViCS ETG, with correlation coefficients of r$_{P}$=0.42 and 0.58 respectively, both with $<1\%$ probability of chance occurrence. The difference in dust mass between the samples is highlighted by the ranges exhibited: 4.1$\le$log(M$_{d}$/M$_{\odot}$)$\le$6.7 for HeViCS and 5.5$\le$log(M$_{d}$/M$_{\odot}$)$\le$7.9 for H-ATLAS. Results from stacked data for red galaxies in (\citealp{bourne_2012}, their fig.~16) show a similar trend to that of the H-ATLAS ETG shown in Fig.~\ref{fig:newdustmassplot}. Following a similar KS test as previously, but using a line through the upper edge of the HeViCS detections in Fig.~\ref{fig:newdustmassplot}, we find 197 H-ATLAS detections above the line, out of 771 GAMA ETG. This gives a KS statistic of 0.25, which for these optical sample sizes gives a probability $<0.001\%$ of the Virgo ETG being drawn from the same parent population as the GAMA ETG. In other words the {\it proportion} of dusty ETG is much higher in the GAMA sample than in the Virgo sample.  

\section{Discussion} \label{sec:section6}

\subsection{Mock Catalogues and Detection Limits}

We further test how the two samples of ETG compare in the left plot of 
Fig.~\ref{fig:dustmass} by carrying out Monte Carlo simulations for the 
Virgo galaxies, to account for the sample selection effects of the GAMA 
galaxies. 500 mock catalogues, each with 771 galaxies (as in the GAMA 
ETG sample), are generated by randomly sampling the Virgo ETG (for their 
dust mass, stellar mass and submm fluxes) and placing them at distances 
randomly selected from the GAMA ETG sample. The observed Virgo ETG submm 
fluxes (or non-detections) in 250 and 350 micron Herschel bands are then 
transformed to fluxes at the new distances, taking into account 
K-corrections (as in equation 2 of Dunne et al. 2011). Thus fluxes 
significantly decrease due to the larger distances and slightly 
increase due to K-corrections in the submm. 

For each mock catalogue the number of expected submm detections 
above the H-ATLAS limits (5~sigma at 250$\mu$m and 3~sigma at 
350$\mu$m, as in A13) is calculated and expressed as a percentage 
of the 771 galaxies. The predicted mean and standard deviation is 
then $0.92\pm 0.36\%$ detections, with standard deviation calculated 
from the spread of results amongst the 500 mock catalogues.
For GAMA ETG 188 out of 771 galaxies were actually detected (above those 
H-ATLAS limits, as in Fig.~\ref{fig:dustmass} left). 
This corresponds to $24.4\pm 2.0\%$, taking into account Poisson 
noise in these numbers. 

Therefore the difference between H-ATLAS ETG detections and 
the mock HeViCS detections at H-ATLAS distances, is greater 
than 11~sigma. This corresponds to a negligible probability 
($<10^{-28}$) of the HeViCS sample being consistent with the H-ATLAS sample, 
in terms of their intrinsic submm properties.
Thus the differences in specific dust distributions seen in 
Fig.~\ref{fig:dustmass} is not explained by different dust mass 
detection limits in the two samples.

\subsection{Environments}

\begin{figure}
\begin{center}
\hspace{-0.2in}
 \includegraphics[width=0.5\textwidth]{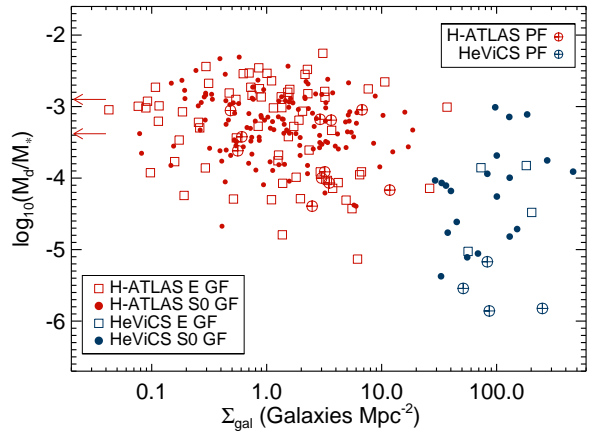}
 \caption{Dust-to-stellar mass ratio derived from MAGPHYS fits plotted against environment surface density for the two ETG samples.  H-ATLAS ETG are plotted in red (Es are open circles and S0s are filled circles) and HeViCS ETG are plotted in blue (Es are open squares and S0s are filled squares). Poor SED fits are marked in encircled crosses (in red or blue based on the sample). Arrows indicate the dust-to-stellar mass ratio for two H-ATLAS S0s at $\Sigma_{\textrm{gal}}=$0.001 gals Mpc$^{-2}$.} \label{fig:newcomparison}
\end{center}
\end{figure}

As it has been shown that MAGPHYS successfully reproduces the modBB results for dust mass, 
we adopt hereafter the parameters ($T_d$, $M_d$ and $M*$) derived with MAGPHYS. This has the advantage of ensuring uniformity of approach and also of providing SF parameters. However, this means that we must exclude from the subsequent analysis the 4 HeViCS ETG with synchrotron radiation and the MAGPHYS poor fits (4 in HeViCS and 11 in H-ATLAS). We note that the multi-waveband photometric datasets used in the MAGPHYS fits are different for the HeViCS and H-ATLAS samples.
We begin by examining the relationship between environment and dust-to-stellar mass ratio in Fig.~\ref{fig:newcomparison}. In spite of the large scatter in this plot, there is a weak overall log-log anticorrelation (r$_{P}$=-0.3), which is influenced by the different sensitivity limits of the two samples.
For H-ATLAS data alone the correlation coefficient decreases to r$_{P}$=-0.2, which is not a significant result, therefore the anticorrelation is not clear within each individual sample. 

For the  H-ATLAS sample, which is less sensitive, we have attempted a stacking analysis to probe the undetected ETG (following the methods of \citet{bourne_2012}), accounting for blending of nearby sources and assuming the flux is spread out over the optical extent of each source \citep{hill_2011}. By stacking at the position of the undetected ETG in the 100 - 500 $\mu$m maps we derive the median value of the flux density in the stack in order to avoid bias from outliers, and we estimate the 1$\sigma$ error on the median from the distribution of values in the stack. From these median stacked flux densities we find a median dust mass of $\sim 4.44 \times $10$^5$M$_{\odot}$, at a median surface density of $\Sigma_{gal}=2.3$ gals Mpc$^{-2}$ and median stellar mass of $\sim2.0 \times 10^{10}$M$_{\odot}$. This is at log$_{10}$(M$_{*}$)=10.3 and log$_{10}$(M$_{d}$/M$_{*}$)=-4.65, which lies amongst the dust detected S0 galaxies in the Virgo cluster in  Fig.~\ref{fig:dustmass} and well above the HeViCS detection limits. In Fig.~\ref{fig:newcomparison} this specific dust mass is comparable to the Virgo detected ETG. However the 90 undetected ETG in the Virgo cluster will lie well below this point in specific dust mass, since they are below the dashed blue lines in Fig.~\ref{fig:dustmass} and the two samples cover similar stellar mass ranges. Conversely, if the detected Virgo ETG were placed at the average distance of the GAMA ETG sample, then only one Virgo ETG would be detected in Fig.~\ref{fig:newcomparison} (VCC1535). It is difficult to quantify the lower limit of the distribution in Fig.~\ref{fig:newcomparison} without further constraints, but it is clear that sensitivity limits are not causing the downward slope seen at the top of the distribution in Fig.~\ref{fig:newcomparison}.

The substantial range of normalised dust mass displayed by the Virgo ETG is a feature which needs to be subjected to more scrutiny, e.g. studying the dust detection rate and normalised dust mass in fast- and slow-rotators, following a suggestion by S13. ATLAS$^{3D}$ is an ongoing survey investigating the kinematic properties of a volume-limited sample of ETG including the Virgo cluster, finding that elliptical galaxies tend to be either slow rotators or rotate faster than lenticulars (based on apparent specific angular momentum). They find that non-rotating ETG tend to be found in highly overdense environments \citep{krajnovic_2011} - their results also indicate that in dense groups and clusters gas accretion is suppressed. Although twenty of the ETG in the HeViCS sample correspond with those studied by ATLAS$^{3D}$ \citep{emsellem_2011}, it is inconclusive from such a comparison whether the rotation speed of each galaxy is related to its respective dust-to-stellar mass ratio, for the current sample. However it has been shown that a galaxy's stellar angular momentum and stellar mass are negatively correlated (e.g. \citealp{emsellem_2011}), and normalised dust mass is negatively correlated with stellar mass (A13). Therefore it can be postulated that dust-to-stellar mass ratio is positively correlated with a galaxy's rotation speed. However, the reverse trend was found in Virgo (S13), probably because in a denser environment the giant slow rotators may be collectors of stripped dust. This suggestion might be investigated in the future with high-resolution, kinematic observations of the cold dust or gas in GAMA/H-ATLAS ETG.

Given the wide range of environments inhabited by the H-ATLAS ETGs, and the large sample size, we can now split H-ATLAS ETG into two subsamples: H-ATLAS-LOW ($\Sigma_{gal}<$1.25 gals Mpc$^{-2}$) and H-ATLAS-HIGH ($\Sigma_{gal}\geq$1.25 gals Mpc$^{-2}$)\footnote{The environment cutoff has been chosen to create subsamples of equivalent size}. Note that such subsampling is not possible for HeViCS ETGs due to the small sample size. We examine average MAGPHYS dust-to-stellar mass ratio for the two subsamples, and subsequently compare with HeViCS dust-to-stellar mass.

H-ATLAS-HIGH has a mean dust-to-stellar mass ratio of log$_{10}$(M$_{d}$/M$_{\ast}$)=-3.35, indicating lower dust fractions on average in comparison to H-ATLAS-LOW (mean log$_{10}$(M$_{d}$/M$_{\ast}$)=-3.19). This may be influenced by the well known property that massive ETG tend to reside in denser environments. In fact H-ATLAS-HIGH has intermediate dust-to-stellar mass ratio between H-ATLAS-LOW and HeViCS (mean log$_{10}$(M$_{d}$/M$_{\ast}$)=-4.55). However, a KS test indicates no significant difference in the distribution of this parameter between the two H-ATLAS subsamples. 

These results for dust parameters can be interpreted in different ways. The subtle decrease in relative dust mass with increasing environment may be a real effect, which is then extended to the dense region of the Virgo cluster, or it may be completely spurious. Our tests using mock catalogues above confirms the real difference between the two samples. More samples at intermediate environment densities and in other cluster environments are needed to test the reality of this trend with environment. 

\subsection{Star Formation Properties}

In A13 UV-optical colour was used as a proxy for SFR in a galaxy. However blue UV-optical colours can also be induced by the presence of a very old stellar population in the galaxy \citep{greggio_1990,horch_1992,bressan_1994} and therefore further investigation of the SFRs in these galaxies is a necessity to confirm the results found thus far. This must be handled with care, as SFRs derived from MAGPHYS are also related to the UV emission; however the inclusion of longer wavelength information and energy balance in SFR calculations gives a better estimation than a simple proxy. Additionally, it should be noted that SFR estimates using FIR data are subject to large errors in galaxies where dust heating is dominated by the diffuse radiation field from an old stellar population \citep{bendo_2012}.

The interpretation of SED fits to ETG data must account for the potential contribution to UV light from old, evolved stars on the horizontal branch (HB). \citet{burstein_1988} showed that populations of stars older than $\sim$10 Gyrs can cause an upturn in the UV flux, due to UV emission from HB stars. Younger populations do not have this component contributing to the integrated UV light. Populations younger than $\sim$3 Gyrs again have excess UV emission but from the young, massive, main sequence stars. Thus all intermediate age populations  ($\sim$3 to 10 Gyr) do not have significant UV upturns in their spectra and hence will show no sign of a UV excess that could be erroneously attributed to the presence of young stars. The GALEX NUV flux is less affected by the UV flux from old, evolved stars than the GALEX FUV band (see \citealp{kaviraj_2007a}, their fig.~1).

A concern with using MAGPHYS for the interpretation of parameters such as SFR, is that they are based on optical, UV and FIR emission and are calibrated only for galaxies where dust heating is primarily contributed to from the young and old stellar population. However, if the photons heating the dust come primarily from UV emission from an old stellar population, then the results obtained with MAGPHYS are less reliable, because of the uncertainty in UV contribution from old stars. This is a known effect, which is partly incorporated into the models of \citet{bruzual_2003} used in MAGPHYS, through the (albeit uncertain) inclusion of some hot evolved stars. \citet{salim_2007} (and references therein) tested nearby elliptical galaxies, also using the models of \citet{bruzual_2003} and found that those models could account for UV light from old stellar populations.  In addition, eyeballing the fits done here indicates that the UV upturn of the ETG is not a strong effect. 

We can further strengthen our argument that UV emission from the old stellar population is not the driving mechanism for the dust heating by examining the NUV-r colour, in relation to that expected from the UV upturn. UV contamination from old stars leads to colours that are still redder than NUV-r$>$5.0 (from the sample of \citet{kaviraj_2007a}; their fig.~11). Their ETG classified as old (ages$>$1 Gyr from stellar population model fits) all have NUV-r$>$5.0. For example, NUV-r = 5.4 is the colour of the strong UV-upturn galaxy NGC 4552 (see \citealp{yi_2005}). In the H-ATLAS/GAMA sample 
141 out of 184 NUV detected ETG are bluer than NUV-r=5.0. This large fraction of blue NUV-r ETG in the A13 sample indicates that the blue colours cannot be explained by a UV upturn from old, evolved stars alone. Therefore the blueness and large scatter of their NUV-r colours (ranging from $\sim$1.5 to 6.5) indicates that these colour and UV fluxes are dominated by different amounts of recent star formation in these ETG rather than by UV emission from old, evolved stars. 
However, we note that for up to 36$\%$ of the 220 dust detected ETG in A13 (including 43 with NUV-r$>$5.0 plus 36 NUV non-detections) the UV radiation and dust heating could be dominated by old stars, making their MAGPHYS SF rates overestimates. The fractional mass involved in the star formation does not need to be very large in order to strongly influence the NUV-r colours of stellar populations (see review by \citealp{kaviraj_2008}, their fig.~1).

The 33 ETG in the HeViCS sample do not exhibit such blue colours, with the average NUV-r$\sim$5.3. Also \citet{sperello_2013b} has shown that dust-detected HeViCS ETG are not bluer in B-H colour than the undetected ones. These facts should be kept in mind when considering the parameters extracted based on MAGPHYS fits to these galaxies, and may in fact result in star formation rate overestimates for Virgo Cluster ETG.

\begin{figure}
\begin{center}
\hspace{-0.1in}
 \includegraphics[width=0.51\textwidth]{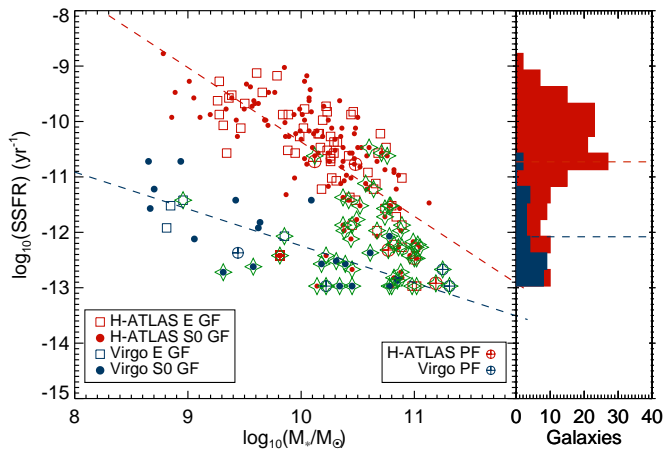}
 \caption{MAGPHYS derived specific star formation rate plotted against stellar mass for the two ETG samples. See Fig.~\ref{fig:newcomparison} for symbols and labels. Green stars overplot points with NUV-r$>$5. Red and blue dashed lines show best linear fits to H-ATLAS and HeViCS respectively. Histograms of these distributions are plotted on the right.} \label{fig:comparison4}
\end{center}
\end{figure}

We examine the sSFR (defined in Section \ref{sec:multiwavelengthfits}) as derived by MAGPHYS below. Fig.~\ref{fig:comparison4} shows sSFR plotted against stellar mass, where a similar trend to Fig.~\ref{fig:dustmass} emerges in the form of an anticorrelation between sSFR and stellar mass of the ETG for the two samples. Regression lines are fit to the two samples, revealing correlation coefficients of r$_{P}=$-0.572 and -0.733 for H-ATLAS and HeViCS samples respectively. These are both significant with much less than $1\%$ probability of occurring by chance. As expected, since derived dust mass and SFR are both influenced by the FIR flux, this corresponds to what was observed for dust-to-stellar mass: H-ATLAS ETG show a weaker correlation for dust-to-stellar mass and sSFR against stellar mass in comparison to HeViCS. Those galaxies with NUV-r$>$5.0 are indicated by green stars in Fig.~\ref{fig:comparison4}. This gives an indication of which points are most likely to be upper limits rather than full SFR measurements in this plot and most of these occur at higher stellar masses. Future work is required to more accurately constrain the low specific star formation rates occurring in massive ETG. 

The side panel of Fig.~\ref{fig:comparison4} shows the distribution of galaxies in sSFR space for the two samples. This quite clearly identifies the higher-on-average sSFR for  H-ATLAS ETG compared with HeViCS ETG, further strengthening the point that H-ATLAS ETG are not only dusty, but actively star forming. The most star forming end of this distribution displays sSFRs which exceed that of our Milky Way Galaxy.  

This anticorrelation between sSFR and stellar mass has previously been observed in both the local and medium-redshift Universe \citep{salim_2007,somerville_2008,firmani_2010}. The study by \citet{salim_2007} took observations of 50,000 SDSS galaxies with a range of morphologies and stellar masses and, after measuring their sSFRs using synthetic population models including dust attenuation, constrained this relation to purely star-forming galaxies as: 
\begin{equation}\label{eq:salim}
\log(sSFR) = -0.35\log M_{\ast} - 6.33.
\end{equation}

\noindent Those star-forming galaxies covered 
$\log(M_{\ast}/M_{\odot}) \sim$8.4 to 11.3 and
$\log(sSFR) \sim$-8.6 to -10.9.
In contrast with this, and over a similar range in stellar mass, the HeViCS ETG display a slightly steeper slope with:
\begin{equation}\label{eq:HeViCS}
\log(sSFR) = -0.59\log M_{\ast} - 6.39\;\;\;  {\rm (NUV\!-\!r=all)}  
\end{equation}
\begin{equation}\label{eq:HeViCS}
\log(sSFR) = -0.39\log M_{\ast} - 8.02\;\;\;  {\rm (NUV\!-\!r<5.0)}
\end{equation}

\noindent and the H-ATLAS ETG produce the steepest gradient of all with: 
\begin{equation}\label{eq:HeViCS}
\log(sSFR) = -1.37\log M_{\ast} + 3.40\;\;\;  {\rm (NUV\!-\!r=all)} 
\end{equation}
\begin{equation}\label{eq:HeViCS}
\log(sSFR) = -0.90\log M_{\ast} - 1.20\;\;\;  {\rm (NUV\!-\!r<5.0)}
\end{equation}

\noindent These differences may be attributed to both the larger sample size and the different galaxy types (presumably dominated by later-type galaxies) making up the relation in \citet{salim_2007}. For the HeViCS and H-ATLAS samples we give above the relations both with and without the green points shown in Fig.~\ref{fig:comparison4} (NUV-r$>$5.0), since those points indicate the least certain associations with ongoin SF. Irrespective of whether those points are included or not, the ETG used in our work span a large range of sSFRs and the relation's gradient is likely to be steepened by the extreme sSFR values. Such a steep relation for the H-ATLAS ETG, coupled with the extreme levels of dust content for lower stellar mass H-ATLAS ETG, is consistent with sSFR downsizing where lower mass ETG harbour star formation in even the local Universe. These low mass ETG also occupy the sparsest environments ($\leq$1 galaxy Mpc$^{-2}$), further strengthening this downsizing theory and in accordance with previous results shown for galaxies in the local Universe \citep{cassata_2007,cooper_2007,wijesinghe_2012}.

What we are seeing in the H-ATLAS ETG sample are 
galaxies like the rejuvenated ones proposed in \citet{Thomas_2010}, (see also 
\citet{Young_2014}).
These rejuvenated ETG have some recent star formation as well as 
old stars and are increasingly common with decreasing galaxy mass.
\citet{Thomas_2010} postulate, from trends seen in SDSS data, with 
environment and chemistry, that these cases, with contributions from 
relatively recent star formation, are decisively influenced by environment.
They support their claim of rejuvination with observations that show
that the more recent star formation is less enhanced in [$\alpha$/Fe], 
and has thus had time to build up more iron from type-Ia supernova, 
for the composition of the later starburst, in contrast to the older stars. 
This trend with environment is similar to the results that we find here. 
Their fig.~8 shows that this is particularly significant for low mass 
ETGs, as we find in Fig.~\ref{fig:comparison4}.

Models for downsizing predict that lower mass galaxies have more extended star formation histories (e.g. \citealp{lucia_2006}). 
If the H-ATLAS sample is divided into low mass (M$_{\ast}<$10$^{10.5}$M$_{\odot}$) and high mass systems (M$_{\ast}\ge$10$^{10.5}$M$_{\odot}$),
it is interesting to note that the lower mass systems exhibit higher average sSFRs over the full redshift range (consistent with results by \citealp{firmani_2010}).
Additionally the higher mass systems show the most similar sSFR to the average sSFR of the HeViCS ETG, indicating greater similarity between these systems than the lower mass ETG have with either of these groupings.

\subsection{Age Properties}

\begin{figure*}
\begin{center}
 \includegraphics[width=0.9\textwidth]{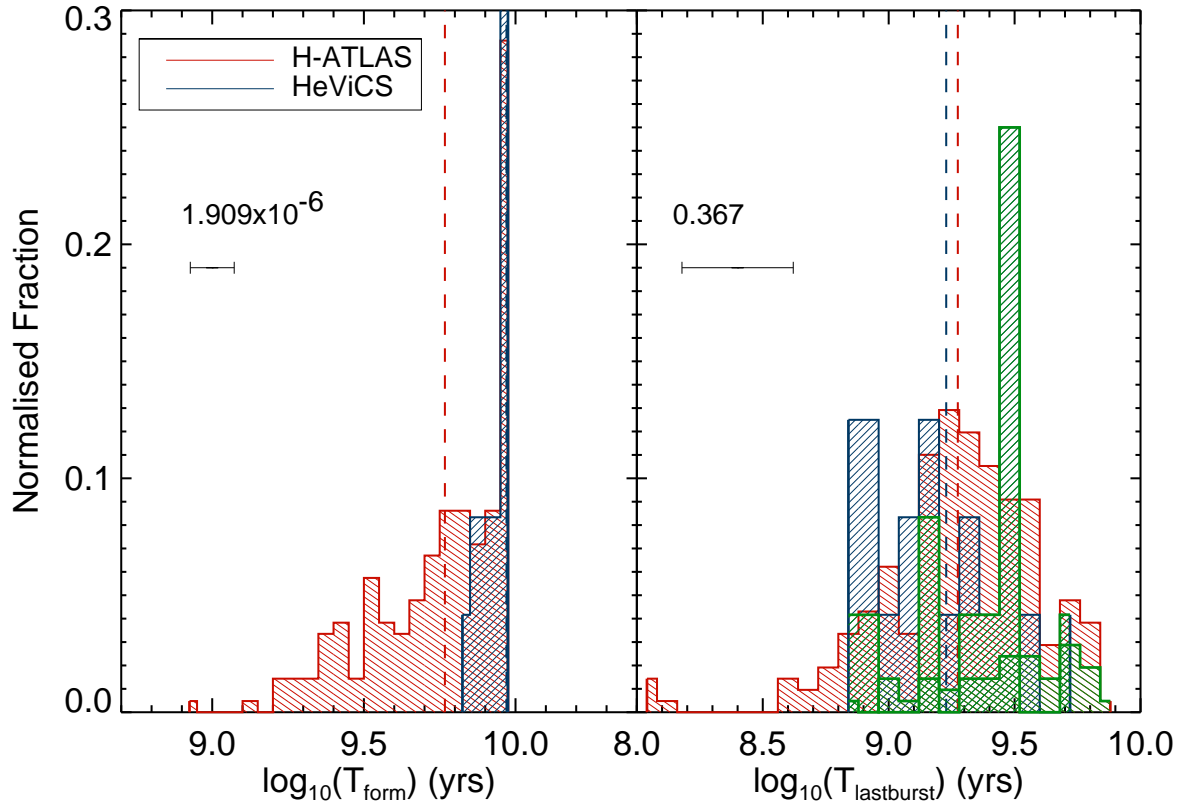}
  \caption{Distributions of $T_\mathrm{form}$ (left) and $T_\mathrm{lastb}$ (right) for H-ATLAS (red) and HeViCS (blue) populations. Numbers in the plots represent KS probabilities of parameter distributions coming from the same parent sample. Dashed lines represent mean values in the samples' respective colours. Error bars represent the mean 1$\sigma$ errors for both samples. For $T_\mathrm{lastb}$ (right) galaxies with NUV-r$>$5.0 are indicated with green histograms.} \label{fig:ages}
\end{center}
\end{figure*}

Ideally, an exploration of the ages of these ETG would begin by using spatially resolved, population synthesis modelling for these systems as a whole. In the case of the H-ATLAS ETG this has not been done yet because current imaging does not have good enough resolution. Therefore we choose to run a pilot study on the ages of these galaxies using MAGPHYS results. Caution must be applied to the use of these results, as they are fully dependent on SPS code used to compute the short-wavelength light produced by stars, which is likewise dependent on the model's choice of metallicity, initial mass function (IMF) and star formation history. In this case, the SPS code is that of \citet{bruzual_2003}, and they adopt a range of exponentially declining SFHs plus bursts and a Chabrier IMF \citep{chabrier_2003}. Here we will only consider relative numbers, and not absolute ages.

We choose to examine two forms of galactic age: the formation timescale ($T_\mathrm{form}$), which is defined as the age of the oldest stars in the galaxy and is a good representative of the age of the galaxy, and the time the last burst of star formation ended ($T_\mathrm{lastb}$). The distributions of both these parameters and means thereof are compared in Fig.~\ref{fig:ages}. Probability results from KS testing the distributions are also included in these figures. The left panel indicates that the two samples have significantly different formation timescales. However, the results for $T_\mathrm{lastb}$ indicate that there is no difference ($<$1 per cent probability of a difference) between the two sets of galaxies. This is an interesting result as it is the first point at which any similarity between the parameters of the two sets of ETG has been found, although $T_\mathrm{lastb}$ is not always well determined. The green histograms plotted for $T_\mathrm{lastb}$ in the right plot in  Fig.~\ref{fig:ages} illustrate where the least certain meaurements are, since these have NUV-r$>$5.0 colours. These timescales are provisional estimates, compared here in a relative way. Further detailed study on the galactic stellar populations is required to determine whether these results are real or simply a result of the assumptions made in the SPS fitting.

\subsection{Additional Considerations}

Based on the results discussed above,
a clear conclusion about the two samples is reaffirmed: that the ETG in each of the samples have differing dust properties, with HeViCS ETG demonstrating consistently low dust levels, whilst the H-ATLAS ETG have significantly higher dust levels which bridge the gap between HeViCS ETG and late-type spirals. The dust-to-stellar mass ratio is shown to be strongly driven by the stellar mass of the galaxy, particularly for the HeViCS systems. This work has made uniform the calculations of stellar and dust mass for two samples; this has served to strengthen the result that there is no overlap between dust-to-stellar mass ratio for fixed stellar mass for ETG from these two samples, covering different environments.

This lack of overlap results partly from different dust detection limits between the two samples; the closer distance to HeViCS ETG and deeper observations result in much lower detectable dust levels in the HeViCS sample. 
From the H-ATLAS Science Demonstration Phase, \citep{rowlands_2012} 
stacked 233 optically selected ETG from GAMA and found that they have average 
dust masses of an order of magnitude less than their H-ATLAS detected ETG.
However, this does not explain why there are
no HeViCS ETG occupying the same regions as the H-ATLAS sub-mm detected ETG. 
This cannot be explained by a detection limit, but may be due to the larger area surveyed by H-ATLAS. Some of the more extreme cases in H-ATLAS could be explained as unusual ETG, but given that all H-ATLAS ETG occupy the top region of Fig.~\ref{fig:dustmass}, this effect is likely explained by the difference in environment. Alternatively, this could be caused by the inability of dust to survive in dense environments due to galaxy-cluster interactions. Hydrodynamical or gravitational interactions that are likely to occur in dense regions may shorten the lifetime of dust, as may hot gas in the hostile intracluster medium.

There are only three HeViCS ETG which demonstrate normalised dust masses on a level with the H-ATLAS ETG (VCC 327, 450 and 571). These all have GOLDMine classifications of S0 and are found in high density regions of Virgo ($\Sigma_{\textrm{gal}}\sim$100-200 gals Mpc$^{-2}$). It may be possible that these galaxies have been recently accreted into the Virgo cluster (e.g. \citealp{kraft_2011}), and have not yet been subjected to the effects of dust stripping and destruction in the intra-cluster medium.

We run a simple test to check whether this may be a possibility. Based on Virgo infall velocities provided by \citet{mould_1999} and assuming a Virgo cluster radius of 2.2 Mpc \citep{mei_2007}, we calculate typical crossing times for these three Virgo ETG of $\sim$0.7, 0.8 and 2.2 Gyr. If we assume ram-pressure stripping is responsible for the majority of dust loss in Virgo ETG, with typical removal timescales of a few $\times$10$^{8}$ yr \citep{takeda_1984,murakami_1999}, then it may be possible to relate these high relative dust levels with recent galactic infall into the Virgo cluster. 
These relatively short crossing times illustrate that the current local density
may not be the same as the time averaged local density experienced by a galaxy in the Virgo cluster. This is in contrast to what happens in the field. This may contribute to the different behaviours with environment shown in Fig.~\ref{fig:betweensamples}.

 Dust masses detected in these galaxies can be used to estimate total (both atomic and molecular) gas mass: a typical gas-to-dust ratio of 100 (e.g. \citealp{parkin_2012})
gives a range of $\sim$1$\times$10$^{7}$-8$\times$10$^{9}$M$_{\odot}$ for the H-ATLAS sample and $\sim$10$^{6}$-5$\times$10$^{8}$M$_{\odot}$ in HeViCS. ATLAS$^{3D}$ estimate molecular gas mass for some of the Virgo galaxies in this sample, finding an upper limit of 10$^{8.59}$M$_{\odot}$ for these particular galaxies \citep{young_2011}, which is consistent with our estimations for total gas mass. Additional results from ATLAS$^{3D}$ find a strong HI detection rate dependence on surface density whereby HI in ETG is preferentially detected outside the Virgo Cluster \citep{serra_2012}; again these results are qualitatively consistent with our findings for the two samples whereby H-ATLAS ETG demonstrate a factor of ten dust and hence gas mass higher than HeViCS ETG in the dense regions of Virgo. However, while we note that dust and gas masses are both low in these Virgo ETG, S13 found that dust and H-I detections in Virgo ETG showed very small overlap, counter to the assumption that dust follows gas mass.
 
 Similar studies run on samples of LTG in the Virgo Cluster have demonstrated appreciably low levels of HI gas compared to LTG in sparser environments. Additionally, lower star formation activity has been identified in these spiral galaxies, and possibly lower dust levels (\citealp{boselli_2006} and references therein). Models indicate that ram pressure stripping, gas compression \citep{byrd_1990,tonnesen_2009} and starvation due to the cluster potential \citep{balogh_2000} are possible causes of these decreased levels of gas and dust in LTG. Theoretically, ETG in the same environment would also be subjected to these same physical mechanisms, resulting in the lower levels of gas and dust currently being observed.

The Herschel Reference Survey (HRS, \citealp{boselli_HRS}) sampled a wider range of galaxy environments than just the Virgo cluster. Although it includes very few luminous ETG other than those in the Virgo cluster, it is still useful to consider where their ETG reside in terms of parameter space, and how this compares to the two samples investigated here. \citet{smith_HRS} find 31 ETG in the HRS parent sample with 250$\mu$m detections: these all have stellar masses at $\gtrsim$10$^{10}$M$_{\odot}$ and modBB fits to the sample reveal a dust mass range of 10$^{5.0-7.1}$M$_{\odot}$ and dust temperatures of 16-32K. Most notably, however, while they find a similar trend for dust-to-stellar mass ratio with stellar mass, their elliptical galaxies are found to present the lowest normalised dust masses. This is not what is seen here, particularly for the H-ATLAS/GAMA sample. It should be noted that the HRS sample only contains seven sub-mm detected elliptical galaxies, and therefore this result may be due to poor statistics. The majority ($\sim$68$\%$) of the HRS ETG reside within the Virgo cluster, which explains the similar dust mass range to that of the HeViCS survey; in fact \citet{smith_HRS} explicitly state that there is overlap between their ETG and those of S13. Therefore we choose not to include a further study with HRS ETG.

In summary we find tentative evidence that specific dust mass depends broadly on environment, however, more galaxy samples at intermediate environment density are needed to confirm such a trend.

\section{Conclusions}

This work has compared H-ATLAS sub-mm detected ETG to HeViCS (Virgo Cluster) sub-mm detected ETG. This was a strongly motivated study, as multiple $Herschel$ works have revealed different levels of dust in different samples of ETG (\citealp{skibba_2011,smith_2011,rowlands_2012}; S13; A13). It has been unclear thus far whether these differences are simply due to different sample statistics and/or selection effects, or whether they are real differences which are a result of the different samples observing different types of ETG.

Two samples were selected for this study: the A13 H-ATLAS sample of 220 ETG and 33 ETG from the HeViCS S13 sample, both with M$_r<-17.4$mag. With the aid of consistent calculations for nearest neighbour density, and MAGPHYS panchromatic SED fitting to the multi-wavelength data, we were able to objectively quantify the true differences in the properties of these ETG. The results of this study are summarised here below.

\begin{description}
\item{(i) Nearest neighbour surface densities revealed true differences in the type of environment in which these ETG reside. H-ATLAS ETG are in isolated environments, spanning $\lesssim$0.1-10 galaxies Mpc$^{-2}$, whereas HeViCS ETG are dominated by the cluster environment ($\sim$25-500 galaxies Mpc$^{-2}$). These results are also true for undetected ETG in each sample, with only a trace overlap in density between samples observed at $\sim$20-100 galaxies Mpc$^{-2}$. 
We find that sub-mm detected ETG in H-ATLAS reside in sparser environments than undetected ETG.
}
\item{(ii) ModBB fits from A13 and S13 reveal different ranges of dust-to-stellar mass ratio, with H-ATLAS ETG demonstrating higher M$_{d}$/M$_{\ast}$ at fixed stellar mass. We prove that this is not a selection effect by carrying out a KS test and by Monte Carlo simulations of Virgo ETG at GAMA ETG distance.  Both these tests confirm that the samples have $<0.1$\% probability of having been drawn from the same specific dust mass distribution. 
}
\item{(iii) MAGPHYS results indicate that it is sometimes difficult to accurately constrain the cold dust temperature, but similar results for dust mass 
were obtained using MAGPHYS and ModBB fits, in spite of these uncertainties.
ModBB fits appear to give higher dust mass of a galaxy in some cases - this may be because the smaller dust grains which emit at higher temperatures are not accounted for in the ModBB fits.}
\item{(iv) MAGPHYS stellar masses including UKIDSS fluxes are lower than those produced by \citet{zibetti_2009} (HeViCS) and the GAMA team \citep{taylor_2011} (H-ATLAS). Both sets of stellar masses indicate that H-ATLAS ETG are more massive on average than HeViCS ETG. }
\item{(v) Correlations are found between dust mass and stellar mass for both H-ATLAS (r$_{P}$=0.42) and HeViCS (r$_{P}$=0.58) ETG. Additionally anticorrelations are found between dust-to-stellar mass ratio and stellar mass, although the trend is shifted upwards (to higher normalised dust mass) for H-ATLAS. Most of this anticorrelation is due to the dust detection limits. However it remains to be understood why there is a lack of massive ETG in Virgo with high dust-to-stellar mass ratios. Investigating dust-to-stellar mass ratio as a function of nearest neighbour density reveals another correlation between the two properties, where both H-ATLAS and HeViCS ETG sit on the same trendline. This is an indicator that levels of dust mass in ETG are affected by their environments.}
\item{(vi) Examinations of sSFR reveal that dust mass is indicative of ongoing star formation in many of these galaxies, but is not directly related, as evidenced by different trends in specific dust mass and sSFR plots with stellar mass. It appears that there is very little (if any) ongoing star formation in the HeViCS ETG, but quite the opposite is true for a large proportion of the H-ATLAS sample, with the highest sSFRs on par with that of spiral galaxies.}
\item{
(vii) The massive ETG in H-ATLAS have similar sSFRs to the HeViCS ETG.}
\end{description}

\section*{Acknowledgments}

We would like to acknowledge and thank Elisabete da Cunha for her contribution of modified MAGPHYS libraries. 
NKA acknowledges the support of the Science and Technology Facilities Council. 
LD, RJI and SJM acknowledge support from the European Research Council Advanced Grant COSMICISM. IDL gratefully acknowledges the support of the Flemish Fund
for Scientific Research (FWO-Vlaanderen). KR acknowledges support from the European Research Council Starting Grant SEDmorph (P.I. V.~Wild). The \textit{Herschel}-ATLAS is a project with \textit{Herschel}; which is an ESA space observatory with science instruments provided by European-led Principal Investigator consortia and with important participation from NASA. The H-ATLAS website is http://www.h-atlas.org/. GAMA is a joint European-Australasian project base
d around a spectroscopic campaign using the Anglo-Australian Telescope. The GAMA input catalogue is based on data taken from the Sloan Digital Sky Survey and UKIRT Infrared Deep Sky Survey. Complementary imaging of the GAMA regions is being obtained by a number of independent survey programs including GALEX MIS, VST KIDS, VISTA VIKING, WISE, \textit{Herschel}-ATLAS, GMRT and ASKAP providing UV to radio coverage. GAMA is funded by the STFC (UK), the ARC (Australia), the AAO, and the participating institutions. The GAMA website is http://www.gama-survey.org/.
We thank Gianfranco De Zotti and Michal Michalowski for helpful comments on an earlier draft of this paper. Thanks to the anonymous referee for useful suggestions that improved the paper.

\bibliographystyle{mn2e}
	\bibliography{EnvPaper_astroph.bib}

\appendix

\section{WISE photometry} \label{app:wise}

The Wide-field Infrared Survey Explorer (WISE) has a beam size of $12.0 ''$ (in the W4 band) as its coarsest resolution. This means most of the galaxies in the Virgo cluster are extended objects in all WISE bands. Magnitudes from the WISE Science Archive \footnote{Science Archive: http://irsa.ipac.caltech.edu} are based on aperture photometry using an elliptical aperture for each galaxy. Foreground stars are not removed, hence diluting the flux measurements. We therefore opted to compute asymptotic fluxes for all galaxies in our sample, taking special care of the contamination by bright foreground stars. 

The images for each galaxy were retrieved from the WISE Science Archive. We choose sufficiently large cut-outs from the All-Sky Atlas in order to cover both the galaxy and its close surroundings. 

The local background of the galaxy is estimated from $10$ boxes of $50 \times 50$ pixels. Three values are computed from these boxes: $1)$ a single background value. A sigma clipped mean was derived from all pixels using a $5\sigma$ rejection threshold and iterations until convergence. The background value was then subtracted from the image. $2)$ A pixel-to-pixel background error. This is the mean of the standard deviations on the pixel values in each  of the boxes. $3)$ A large scale background variation error. This is the standard deviation of the mean values of each of the boxes.

\begin{figure}
	\resizebox{\hsize}{!}{\includegraphics{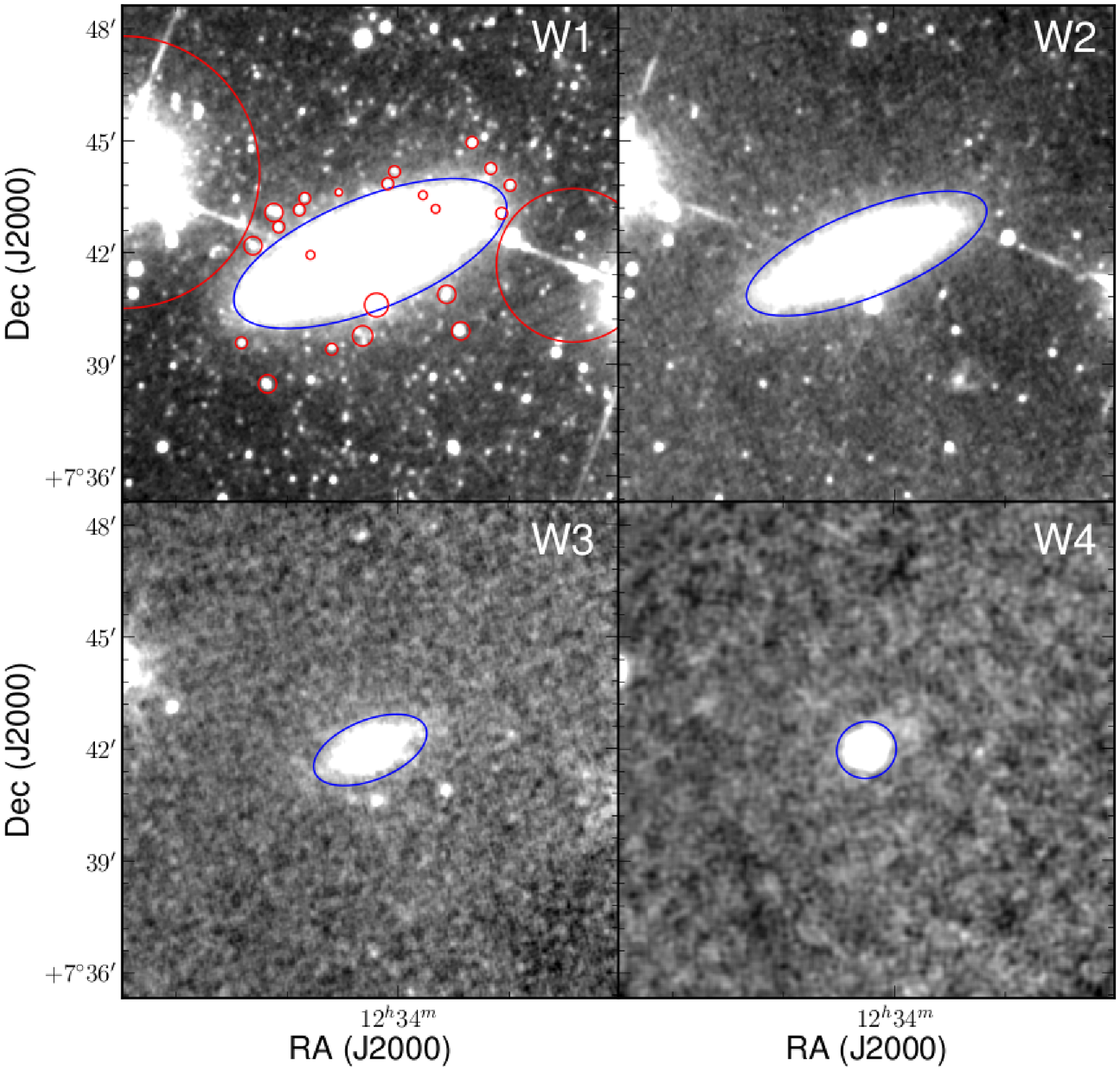}}
	\caption{Example galaxy VCC 1535 in all four WISE bands. In each frame the blue ellipse is the parent aperture for that band from which concentric ellipses were derived. In the W1 frame (top left), the masked foreground stars are indicated in red circles. These pixels were masked in all bands.}
	\label{fig:regions}
\end{figure}

We identify the brightest stars and mask their corresponding pixels so they will be ignored in the further analysis. An elliptical aperture is then determined, following the galaxy's apparent shape on the sky. The angle and ellipticity of this parent aperture will serve as a base for constructing the growth curve.
To construct the growth curve for each galaxy and each band, concentric elliptical annuli are created based on the parent aperture. Starting from the center of the galaxy, each next aperture is a fraction $\Delta a$ in major axis larger. We set $\Delta a = a/20$, where $a$ is the major axis of the parent aperture. The total flux inside the elliptical shell is estimated by multiplying the median value of all pixels inside the  shell with the total number of pixels. In the number of pixels we also include the amount of masked pixels. This extra step filters out any remaining contribution of foreground stars, while approximating the flux of the galaxy in that shell. As soon as the calculated flux in a shell falls below the background level, the iteration is stopped.

\begin{figure}
	\resizebox{\hsize}{!}{\includegraphics{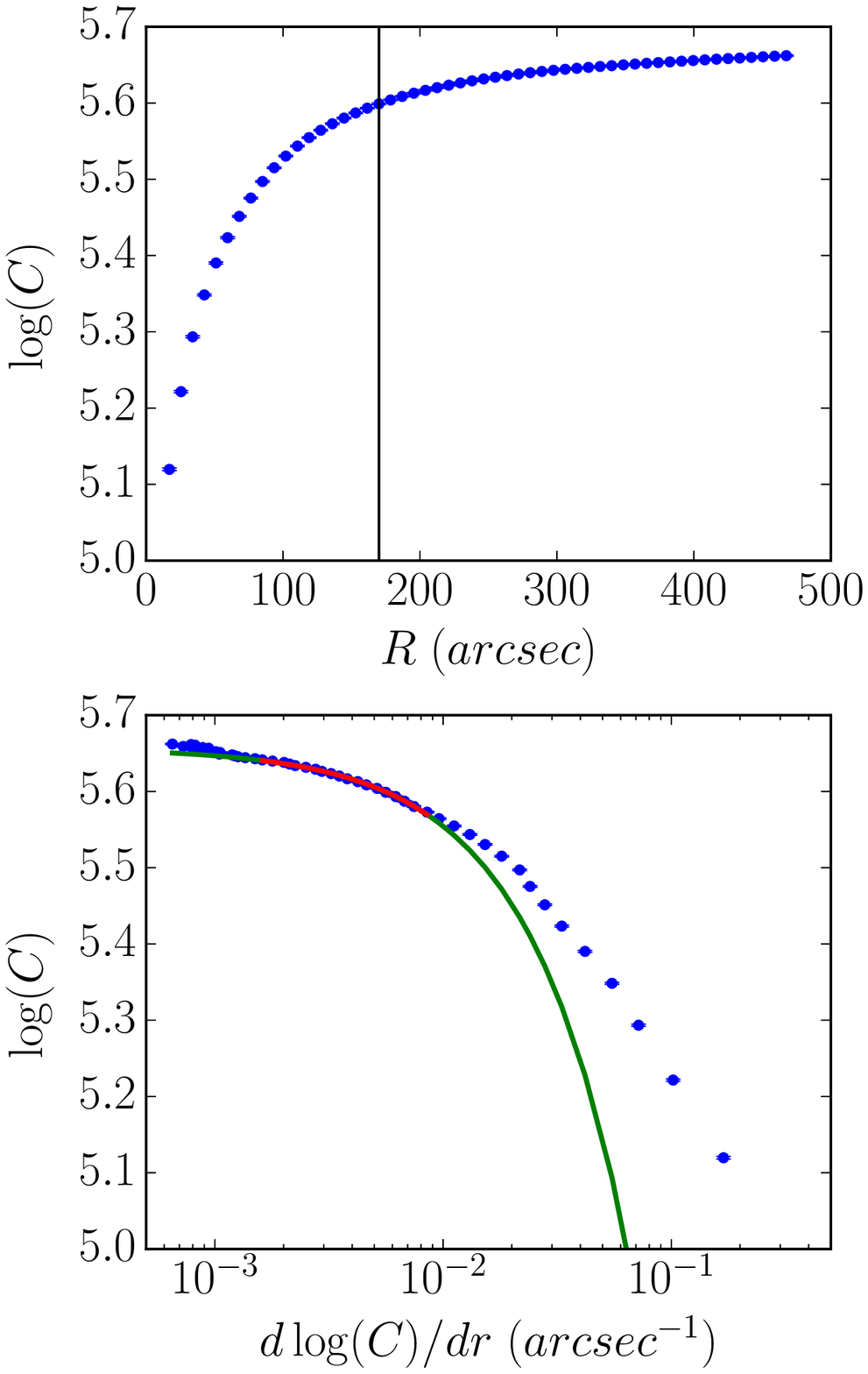}}
	\caption{\textbf{Top}: Growth curve for the W1 band of VCC 1535, showing the log of the cumulative counts for each elliptical shell. The black vertical line indicates the major axis of the parent aperture. \textbf{Bottom}: Cumulative counts as a function of the gradient of the growth curve. The red line is a linear fit (displayed here on log-linear scale) to the blue points in that interval. The green line is the extrapolation of this linear fit to all of the points.}
	\label{fig:curve}
\end{figure}

The growth curve is now plotted as the cumulative counts versus the distance to the center of the galaxy. Following \citet{MunozMateos2009}, the gradient of the growth curve around the edge of the galaxy is calculated. Fig.~\ref{fig:curve} (lower panel) shows the cumulative counts as a function of this gradient. A linear fit to these points is performed and the intercept with the y-axis is calculated as the asymptotic flux of the galaxy in that particular band. As the gradient changes rapidly and is non-linear inside the parent annulus of the galaxy, these points are not considered in the fit. The outer points of the growth curve are not considered in the fit either, as they are usually dominated by background variations or unmasked foreground stars.

Four sources of uncertainty on the calculated fluxes are considered. \begin{inparaenum}[\itshape a\upshape)]
\item The Poisson noise, determined as the square root of the asymptotic flux. A multiplicative factor of $2$ for W1,W2 and W3 and $4$ for W4 must be applied to incorporate the correlated pixel noise, according to the Explanatory Supplement to the WISE Preliminary Data Release Products \footnote{\label{WISEmanual}http://wise2.ipac.caltech.edu/docs/release/prelim/expsup/\\wise\_prelrel\_toc.html}, section II.3.i. \item The pixel-to-pixel background error as described above. \item Large scale background variations, also described above. \item A calibration uncertainty. Following the recommendations of \citet{Jarrett2011}, we use $2.4\%$, $2.8\%$, $4.5\%$ and $5.7\%$ as the calibration error in W1, W2, W3 and W4, respectively.\end{inparaenum}

The obtained fluxes and their respective errors have to be converted to Jy to be of physical meaning. Section II.3.f of the WISE photometry manual \footnote{See footnote \ref{WISEmanual}}, provides the following conversion factors: $1.9350\times 10^{-6}$ Jy/DN
, $2.7048\times 10^{-6}$ Jy/DN, $2.9045\times 10^{-6}$ Jy/DN and $5.2269\times 10^{-6}$ Jy/DN for W1, W2, W3 and W4, respectively.

\citet{Jarrett2013} advises three corrections to the WISE flux of extended sources. One of them is a colour correction, which we do not apply here as our fitting routine takes the filter response into account. The second correction stems from a calibration discrepancy between blue stars and red galaxies, described by \citet{WISE} and \citet{Jarrett2011}. They advise a multiplicative factor of $0.92$ for the W4 flux of all spiral and disk galaxies. As we are dealing with early type galaxies in our sample, we have no need of such a correction. This leaves us with the third correction, which is an aperture correction due to the fact that the absolution calibration for WISE was done using PSF profile fitting. We apply a correction of $0.03$ mag, $0.04$ mag, $0.03$ mag and $-0.03$ mag for W1, W2, W3 and W4, respectively.

\begin{figure}
	\resizebox{\hsize}{!}{\includegraphics{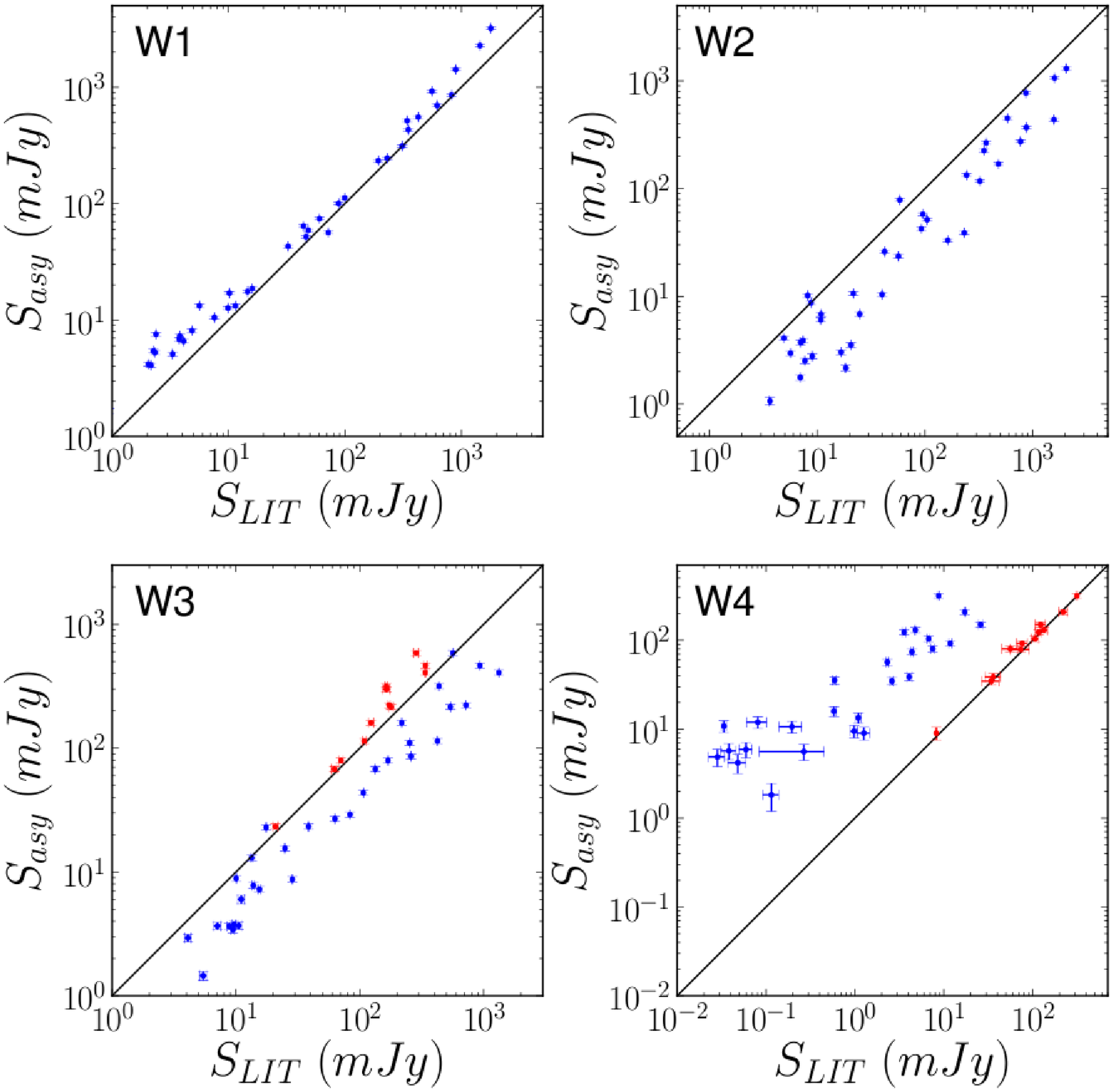}}
	\caption{Comparison of the integrated WISE fluxes as derived by our method (asymtotic fluxes) and the archival fluxes (blue points) measured from elliptical annuli. Red points are independent photometric measurements in W3 and W4 from \citet{Ciesla2014} for a subset of our sample. The black line is the $1:1$ relation.}
	\label{fig:compare}
\end{figure}

We verify our measurements by comparing them with the archival and literature WISE fluxes, determined by flux measurements of elliptical apertures (see Fig.~\ref{fig:compare}). The slopes of the data points in each of the panels are roughly parallel to the $1:1$ relation (black line). There is, however, an offset visible in all bands when comparing to the archival fluxes (blue points). Interestingly, we find systematically higher fluxes in the W1 and W4 band, while our measurements yield lower fluxes in the W2 and W3 bands.

\citet{Ciesla2014} measured the W3 and W4 fluxes of a small subset of our sample as part of the Herschel Reference Survey \citep{boselli_2010}. The corresponding galaxies are of the largest in our sample and include the $4$ radio galaxies. Although the measurements were also done using aperture photometry, much more care was taken in the choice of the apertures than the automated WISE pipeline. Furthermore, the contribution of foreground stars is less significant in these two bands. Our agreement in both W3 and W4 with \citet{Ciesla2014} boosts confidence in our method and recommends independent flux measurements over archival WISE fluxes when it comes to extended sources. Our resultant fluxes for these WISE bands are displayed fully in Table~\ref{tab:fluxes}.

\begin{table*}
\caption{WISE photometry for the HeViCS sample. All measurements are in mJy.}
\label{tab:fluxes}
\centering     
\begin{tabular}{ccccc}
\hline
\hline
VCC	&	$F_\mathrm{W1}$			&	$F_\mathrm{W2}$			&	$F_\mathrm{W3}$			&	$F_\mathrm{W4}$	\\
\hline	
94	&	42.8	$\pm$	1.1	&	23.54	$\pm$	0.78	&	22.9	$\pm$	1.1	&	--	\\
220	&	100.3	$\pm$	2.6	&	51.6	$\pm$	1.6	&	26.9	$\pm$	1.3	&	13.4	$\pm$	1.8	\\
270	&	8.12	$\pm$	0.27	&	3.54	$\pm$	0.18	&	8.89	$\pm$	0.49	&	12	$\pm$	1.8	\\
278	&	6.58	$\pm$	0.23	&	3.73	$\pm$	0.18	&	--	&	--	\\
312	&	51.8	$\pm$	1.4	&	26	$\pm$	0.86	&	13.02	$\pm$	0.68	&	--	\\
327	&	4.18	$\pm$	0.17	&	2.52	$\pm$	0.14	&	3.77	$\pm$	0.24	&	4.9	$\pm$	1.1	\\
345	&	515	$\pm$	13	&	266.6	$\pm$	7.9	&	110.2	$\pm$	5.2	&	74.1	$\pm$	5.7	\\
408	&	233.5	$\pm$	5.9	&	117.6	$\pm$	3.5	&	85.8	$\pm$	4.1	&	56.8	$\pm$	4.7	\\
411	&	18.61	$\pm$	0.54	&	10.4	$\pm$	0.4	&	--	&	--	\\
450	&	5.3	$\pm$	0.2	&	2.16	$\pm$	0.13	&	8.72	$\pm$	0.47	&	10.8	$\pm$	1.6	\\
462	&	64	$\pm$	1.7	&	42.4	$\pm$	1.3	&	29.1	$\pm$	1.4	&	16	$\pm$	2	\\
482	&	13.2	$\pm$	0.4	&	8.72	$\pm$	0.34	&	3.67	$\pm$	0.25	&	--	\\
571	&	5.4	$\pm$	0.2	&	2.97	$\pm$	0.16	&	3.43	$\pm$	0.23	&	5.7	$\pm$	1.2	\\
672	&	17.54	$\pm$	0.51	&	10.6	$\pm$	0.4	&	7.78	$\pm$	0.43	&	10.6	$\pm$	1.6	\\
685	&	245	$\pm$	6.2	&	134	$\pm$	4	&	67.7	$\pm$	3.2	&	34.7	$\pm$	3.3	\\
758	&	74.3	$\pm$	1.9	&	39	$\pm$	1.2	&	--	&	--	\\
763	&	1417	$\pm$	35	&	775	$\pm$	23	&	316	$\pm$	15	&	92.4	$\pm$	6.8	\\
781	&	7.29	$\pm$	0.25	&	3.88	$\pm$	0.19	&	2.9	$\pm$	0.2	&	--	\\
881	&	1902	$\pm$	47	&	1026	$\pm$	30	&	298	$\pm$	14	&	79	$\pm$	6	\\
951	&	13.3	$\pm$	0.4	&	6.06	$\pm$	0.26	&	3.69	$\pm$	0.24	&	--	\\
1003	&	697	$\pm$	17	&	370	$\pm$	11	&	222	$\pm$	10	&	103.9	$\pm$	7.5	\\
1030	&	311.8	$\pm$	7.9	&	169.5	$\pm$	5.1	&	114.1	$\pm$	5.4	&	123.3	$\pm$	8.6	\\
1154	&	553	$\pm$	14	&	274.5	$\pm$	8.2	&	215	$\pm$	10	&	130	$\pm$	9	\\
1226	&	3203	$\pm$	80	&	1307	$\pm$	38	&	587	$\pm$	28	&	150	$\pm$	10	\\
1250	&	59	$\pm$	1.5	&	33.1	$\pm$	1.1	&	43.6	$\pm$	2.1	&	35.4	$\pm$	3.4	\\
1253	&	430	$\pm$	11	&	225.7	$\pm$	6.7	&	79.4	$\pm$	3.8	&	38.7	$\pm$	3.6	\\
1316	&	2275	$\pm$	57	&	1067	$\pm$	31	&	464	$\pm$	22	&	208	$\pm$	13	\\
1535	&	854	$\pm$	21	&	437	$\pm$	13	&	407	$\pm$	19	&	317	$\pm$	20	\\
1614	&	10.5	$\pm$	0.33	&	6.82	$\pm$	0.28	&	3.65	$\pm$	0.24	&	1.82	$\pm$	0.63	\\
1619	&	112	$\pm$	2.9	&	57.9	$\pm$	1.8	&	23.3	$\pm$	1.2	&	9	$\pm$	1.5	\\
1632	&	922	$\pm$	23	&	451	$\pm$	13	&	160.5	$\pm$	7.6	&	80.2	$\pm$	6.2	\\
486	&	17	$\pm$	0.5	&	10.19	$\pm$	0.39	&	2.9	$\pm$	0.2	&	5.6	$\pm$	1.1	\\
1327	&	56.5	$\pm$	1.5	&	78.6	$\pm$	2.4	&	15.53	$\pm$	0.79	&	9.5	$\pm$	1.5	\\
\hline 
\end{tabular}
\end{table*}

\end{document}